\documentclass[12pt, draftclsnofoot, onecolumn]{IEEEtran}

\usepackage{cite}

\usepackage{amsmath}

\usepackage{algorithmic}

\usepackage{array}

\ifCLASSOPTIONcompsoc
  \usepackage[caption=false,font=normalsize,labelfont=sf,textfont=sf]{subfig}
\else
  \usepackage[caption=false,font=footnotesize]{subfig}
\fi

\usepackage{url}
\usepackage{graphicx,amsmath,amssymb,amsfonts}
\usepackage{algorithmic,algorithm}

\usepackage{xcolor}

\newtheorem{remark}{\textbf{Remark}}
\newtheorem{theorem}{\textbf{Theorem}}

\newtheorem{lemma}{\textbf{Lemma}}

\newtheorem{corollary}{\textbf{Corollary}}

\newtheorem{proposition}{\textbf{Proposition}}
\newtheorem{definition}{\textbf{Definition}}

\renewcommand{\IEEEQED}{\IEEEQEDopen}

\makeatletter
\newcommand{\rmnum}[1]{\romannumeral #1}
\newcommand{\Rmnum}[1]{\expandafter\@slowromancap\romannumeral #1@}
\makeatother

\usepackage{setspace}

\begin{document}
\bstctlcite{ref:BSTcontrol}

\title{Resource Allocation for Multi-Cell \\ IRS-Aided NOMA Networks}

\author{Wanli~Ni,~
        Xiao~Liu,~
		Yuanwei~Liu,~
		Hui~Tian,~
		and~Yue~Chen
\thanks{Part of this work has been accepted by the IEEE GLOBECOM Workshop on Advanced Technology for 5G Plus, Taipei, Taiwan, Dec. 2020 \cite{Ni2020Joint}.}
\thanks{W. Ni and H. Tian are with the State Key Laboratory of Networking and Switching Technology, Beijing University of Posts and Telecommunications, Beijing, China (e-mail: charleswall@bupt.edu.cn; tianhui@bupt.edu.cn).}
\thanks{X. Liu, Y. Liu, and Y. Chen are with the School of Electronic Engineering and Computer Science, Queen Mary University of London, London, UK (e-mail: x.liu@qmul.ac.uk; yuanwei.liu@qmul.ac.uk; yue.chen@qmul.ac.uk).}
}


\maketitle

\begin{abstract}
	This paper proposes a novel framework of resource allocation in multi-cell intelligent reflecting surface (IRS) aided non-orthogonal multiple access (NOMA) networks, where an IRS is deployed to enhance the wireless service.
	The problem of joint user association, subchannel assignment, power allocation, phase shifts design, and decoding order determination is formulated for maximizing the achievable sum rate.
	The challenging mixed-integer non-linear problem is decomposed into an optimization subproblem (P1) with continuous variables and a matching subproblem (P2) with integer variables.
	In an effort to tackle the non-convex optimization problem (P1), iterative algorithms are proposed for allocating transmission power, designing reflection matrix, and determining decoding order by invoking relaxation methods such as convex upper bound substitution, successive convex approximation, and semidefinite relaxation.
	In terms of the combinational problem (P2), swap matching-based algorithms are developed for achieving a two-sided exchange-stable state among users, BSs and subchannels.
	Numerical results demonstrate that:
	\rmnum{1}) the sum rate of multi-cell NOMA networks is capable of being increased by 35\% with the aid of the IRS;
	\rmnum{2}) the proposed algorithms for multi-cell IRS-aided NOMA networks can enjoy 22\% higher energy efficiency than conventional NOMA counterparts;
	\rmnum{3}) the trade-off between spectrum efficiency and coverage area can be tuned by judiciously selecting the location of the IRS.
\end{abstract}


\begin{spacing}{1.4}
\section{Introduction}
\IEEEPARstart{T}{he} ever-increasing deployment of wireless devices have placed unprecedented requirements on spectrum, energy and cost efficiency for the forthcoming 5G/beyond networks.
By modifying the amplitude and phase of reflective signals, the software-controlled intelligent reflecting surfaces (IRSs) can reconfigure the wireless channels between transmitters and receivers \cite{Wu2020Towards}.
This remarkable feature of IRSs can be utilized to enhance the performance of wireless communication networks from various aspects such as coverage extension, secrecy improvement, and fairness guarantee.
Compared to the conventional active relays supporting massive multiple-input multiple-output (MIMO) \cite{Lu2014MIMO} or millimeter wave (mm-Wave) communication \cite{Roh2014mmWave}, 
decode or amplify are not requested in the IRS-aided wireless networks due to the reason that the IRS is equipped with a large number of passive reflecting elements.
Thus, both hardware cost and energy consumption of the IRS-aided wireless networks are lower than the conventional amplify-and-forward (AF) and decode-and-forward (DF) schemes \cite{Wu2019IRS}.
Meanwhile, IRSs are capable of operating in a full-duplex and noise-free manner, which leads to improved spectrum efficiency.
{Furthermore, by virtue of its scattering feature, IRSs can be leveraged to assist multi-cell communications by redirecting the incident signals toward one or multiple desired directions \cite{Huang2020Holographic}.}
Given the aforementioned advantages of IRSs, they are recognized as promising candidates for signal enhancement, energy saving and cost reduction in the next-generation wireless networks.

Recently, by simultaneously transmitting the superimposed signal to multiple users on the same frequency, non-orthogonal multiple access (NOMA) scheme has been deemed as a promising technique for enhancing the network performance in terms of throughput and connectivity \cite{Liu2020NOMA}.
In sharp contrast to the conventional orthogonal multiple access (OMA) schemes, the signal for different user is distinguished in the power domain, and the successive interference cancellation (SIC) approach is adopted to decode their desired informations at the receivers \cite{Liu2017NOMA}.
Therefore, it is of great significance for NOMA networks to jointly optimize the power allocation and decoding order to improve the spectrum and energy efficiency \cite{Ding2017Application}, as well as reduce interference.
More particularly, for the multi-cell NOMA networks with large-scale devices, the co-channel interference makes the resource allocation problem among base stations (BSs) coupled
and correlated \cite{Cui2018QoE}, which leads to a challenging optimization problem.
Given these challenges, it is particularly important to jointly design user scheduling and resource allocation for performance improvement in the multi-cell NOMA networks.

Inspired by the advantages of both IRSs and NOMA, it is valuable and imperative to integrate them together to further improve the spectrum and energy efficiency, coverage and connectivity, due to the following profits and reasons:
\begin{itemize}
	\item Firstly, the interference can be suppressed by applying IRSs into multi-cell NOMA networks and properly designing the reflection matrix of IRSs. 
	The desired signal can be enhanced by IRSs, which leads to improved system throughput and reduced energy consumption.
	\item {Secondly, for the cell-edge NOMA users that suffer high signal attenuation,} IRSs can be deployed to passively relay the intended signal in a low-cost way, and thus the coverage of NOMA networks is extended.
	Namely, IRSs are beneficial to provide better service for these cell-edge users with poor signal strength.
	\item Thirdly, the SIC decoding performance will be significantly degraded when users' original	channels are not aligned, then the decoding order of users can be effectively tuned by adjusting IRSs to reconfigure the propagation environment. Therefore, IRSs are also profitable to optimize the user pairing and connectivity.
\end{itemize}

\subsection{Related Works}

\subsubsection{Resource Allocation in NOMA Networks}
In order to avoid the exponential complexity brought by the interaction between inter-cell interference and SIC decoding, many research contributions focus on the simplified single-cell NOMA networks \cite{Lei2016Power,Fang2016Energy,Liu2018Dynamic,Cui2018Optimal,Ding2016Impact,Cui2018Outage}.
With the aim of improving the energy efficiency, 
Fang \textit{et al.} \cite{Fang2016Energy} utilized the difference of convex (DC) programming to solve the power allocation problem, and a suboptimal matching algorithm was developed for subchannel assignment.
To strike a balance between the system throughput and user fairness, Liu \textit{et al.} \cite{Liu2018Dynamic} proposed a dynamic power allocation algorithm to maximize the weighted sum-rate by taking into account of the difference of user channel states.
By applying NOMA into the mm-Wave network, Cui \textit{et al.} \cite{Cui2018Optimal} first leveraged the branch and bound method to find a global optimal solution for power allocation, and then a low-complexity algorithm was developed with the aid of successive convex approximation (SCA) method.
The impact of user pairing on sum rate and outage probability was investigated in \cite{Ding2016Impact}, where both numerical and analytical results demonstrated that NOMA can provide better performance than conventional OMA by exploiting the distinctive channel conditions among users.
Subject to the reliability constraints in the MIMO-NOMA network, the authors of \cite{Cui2018Outage} proposed a joint power allocation and receive beamforming algorithm to maximize the fairness-based system utility under imperfect channel state information (CSI) feedback.

Due to the coupled resource allocation and user pairing problem in the multi-cell NOMA networks \cite{You2018Resource,Liu2017Grouping,Fu2017Distributed,Lei2019Load,Cui2018QoE,Zhao2017Spectrum}, it is non-trivial to optimize them jointly.
For the purpose of maximizing the energy efficiency, low-complexity algorithms were developed in \cite{Liu2017Grouping} to solve the resource allocation problem by adopting the matching theory and DC programming.
Furthermore, to reduce the overheads brought by the information exchange among BSs, Fu \textit{et al.} \cite{Fu2017Distributed} designed a fully distributed power control algorithm to minimize the total transmission power at the transmitters while satisfying the data rate requirements of all users.
%
%
%
Taking both user fairness and spectrum efficiency into consideration, Zhao \textit{et al.} \cite{Zhao2017Spectrum} adopted the matching game and SCA methods to iteratively update spectrum allocation and power control results, where a near-optimal solution can be found within a limited number of iterations.

\subsubsection{IRS-Aided Wireless Communication Networks}
The majority of existing research contributions on IRS-aided wireless networks focus on the theoretical analysis \cite{Hou2020Reconfigurable, Zhang2020Capacity, Cheng2020Downlink, Ding2020IRS, Zhang2020Reconfigurable}
and performance optimization in terms of the system throughput \cite{Mu2020Capacity, Zuo2020Resource, Pan2020Multicell, Zuo2020Intelligent},
energy efficiency \cite{Huang2019EE, Liu2020RIS, Wu2020Beamforming, Zheng2020IRS},
and user fairness \cite{Xie2020Max}.
%
By considering the perfect and imperfect SIC decoding of the IRS-aided NOMA network, the authors of \cite{Hou2020Reconfigurable,Zhang2020Capacity,Cheng2020Downlink} derived the closed-form expressions for the outage probability and ergodic rate.
%
%
Due to the hardware limitations of IRSs in practice, the impacts of finite-resolution amplitude and phase shifts on outage probability and achievable data rate were analyzed in \cite{Ding2020IRS} and \cite{Zhang2020Reconfigurable}, respectively.
%
By considering the ergodic and delay-limited capacity for IRS-aided OMA and NOMA networks, Mu \textit{et al.} \cite{Mu2020Capacity} jointly optimized the phase shifts and resource allocation to maximize the average sum rate of all users by invoking the Lagrange duality method.
%
%
%
%
With the objective to minimize the transmission power at the access point, Wu \textit{et al.} \cite{Wu2020Beamforming} proposed both optimal and suboptimal algorithms to design the active and passive beamforming alternately in both the single-user and multi-user cases.
%
%
%
Based on the second-order-cone programming and semidefinite relaxation, Xie \textit{et al.} \cite{Xie2020Max} maximized the received minimal signal-to-interference-plus-noise ratio (SINR) to guarantee user fairness in IRS-aided multiple-input single-output (MISO) networks.

\subsection{Motivations and Contributions}
Inspired by the aforementioned benefits of both IRSs and NOMA, the IRS-aided NOMA transmission scheme can be regarded as an innovative and promising candidate for the next-generation networks.
Although some research contributions on IRS-aided NOMA networks have addressed the challenging transmission power and reflection beamforming optimization problem iteratively, the system models are limited to single-cell and/or single-carrier setups \cite{Hou2020Reconfigurable, Zuo2020Resource, Mu2020Capacity}.
The motivations and challenges of this paper are summarized as follows:
\begin{itemize}
	\item 
	Currently, there is still a paucity of research contributions on investigating the IRS-aided multi-cell NOMA networks with multiple subchannels,
	especially for the user association and resource allocation problem with the mutual SIC decoding constraints and individual quality of service (QoS) constraints.
	\item 
	So far, it is still a challenging issue to maximize the achievable sum rate by jointly designing the transmission power, reflection matrix, and decoding order, while guaranteeing the QoS requirements of all users within the available power budget.
	\item Moreover, the combinational optimization with respect to ( w.r.t.) the user association and subchannel assignment is NP-hard. 
	The complexity of exhaustive search is exponential, and it is non-trivial to obtain an optimal scheme in the polynomial-time complexity. 
\end{itemize}

{
In order to tackle the aforementioned challenges, we study the resource allocation problem in IRS-aided multi-cell NOMA networks to maximize the sum rate, especially the interplay between IRS and NOMA.
Compared to the single-cell NOMA network in \cite{Mu2020Capacity} and \cite{Zuo2020Resource}, the decoding order optimization of the multi-cell NOMA networks becomes more complex even without integrating IRSs into the networks.
Different from \cite{Cui2018QoE} and \cite{Lei2019Load}, the highly coupled variables in multi-cell IRS-aided NOMA networks make the signal processing and performance improvement more complicated, e.g., the co-design of decoding order and reflection matrix is investigated in our work.}
Against the aforementioned background, the main contributions of this paper are summarized as follows:
\begin{enumerate}
	\item 
	We propose a novel framework of resource allocation in the multi-cell NOMA network for enhancing the spectrum efficiency with the aid of a single IRS.
	We formulate the sum-rate maximization problem subject to the SIC decoding conditions, QoS requirements, and maximum power constraints by jointly optimizing the decoding order, transmission power, reflection matrix, user association, and subchannel assignment. We analyze that the formulated problem is a mixed-integer non-linear programming (MINLP) problem, which is NP-hard and is non-trivial to solve directly.
	\item In order to tackle the non-linear optimization problem of joint power allocation, reflection matrix design and decoding order determination, we first adopt relaxation methods such as convex upper bound substitution and SCA to transform the non-convex constraints into convex ones, which can be solved by suboptimal solutions with polynomial time complexity. Afterwards, we invoke the semidefinite relaxation (SDP) and Gaussian randomization methods to handle the rank-one constraint. Finally, the decoding order is obtained according to the combined channel gains arranged in ascending order.
	\item In an effort to solve the three-dimensional (3D) matching problem among users, BSs and subchannels, we first reformulate the decomposed two-dimensional (2D) subproblems into many-to-many (one) matching games that have peer effects but lack substitutability.
	Then, based on the swap operation, we develop two efficient matching algorithms to achieve a two-sided exchange-stable state among the involved players.
	Finally, we analyze the stability, convergence, complexity and optimality of the proposed algorithms from a theoretical perspective.
	\item 
	We demonstrate that the proposed resource allocation algorithms outperform the benchmarks in terms of sum rate and energy efficiency, while NOMA is capable of achieving a better performance than conventional OMA. Additionally, the performance of NOMA networks can be further improved with the aid of the IRS.
\end{enumerate}

\begin{figure} [t!]
	\centering
	\includegraphics[width=3.5 in]{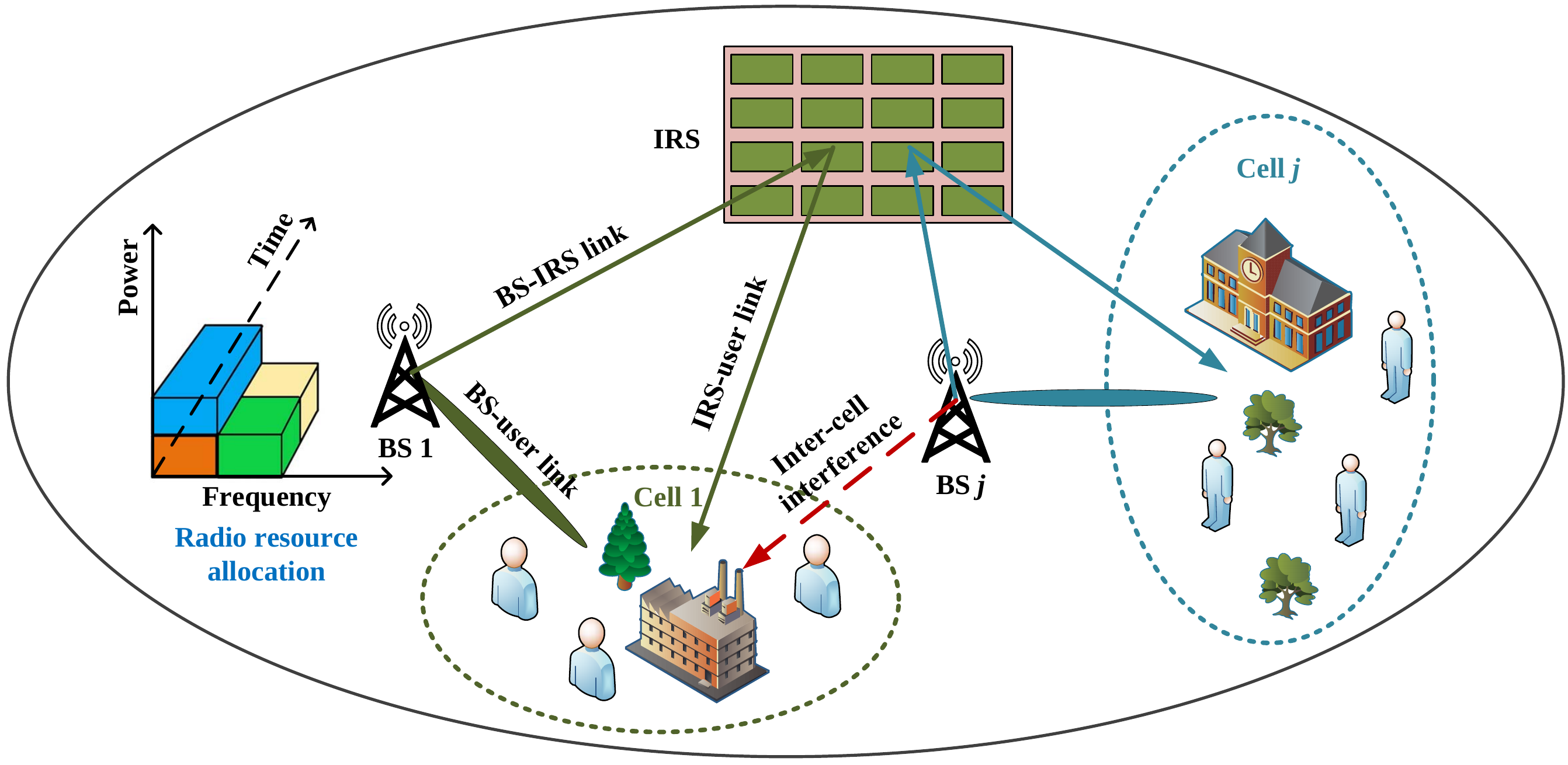}
	\caption{An illustration of the system model for the IRS-aided multi-cell NOMA network, where an IRS with $M$ reflecting elements is deployed to assist the wireless communication from $J$ single-antenna BSs to $I$ single-antenna users.
	}
	\label{system_model}
\end{figure}

The rest of this paper is organized as follows.
First, the system model and problem formulation of the IRS-aided multi-cell NOMA network is given in Section \ref{system}.
Then, the sum-rate maximization problem is solved in Section \ref{optimization} and \ref{matching}.
Finally, numerical simulations are presented in Section \ref{simulation},
which is followed by conclusions in Section \ref{conclusion}.


\section{System Model and Problem Formulation}
\label{system}
\subsection{System Model}
As illustrated in Fig. \ref{system_model}, we consider an IRS-aided multi-cell NOMA transmission scenario\footnote{This paper focuses on the single-antenna case to work on a neat model for providing succinct insights, the algorithms proposed in this paper can be extended to the multi-antenna case as well by exploiting the problem similarity, details of which are omitted here for simplicity.}, where an IRS is deployed for enhancing wireless service from $J$ single-antenna BSs to $I$ single-antenna cellular users, while $\mathcal{I}=\{1,2,\dots, I\}$ and $\mathcal{J}=\{1,2,\dots, J\}$.
It is assumed that each cell is served by one BS, and each cellular user has to be associated with one BS.
The IRS is equipped with $M$ passive reflecting elements, denoted by $\mathcal{M}=\{1,2,\dots, M\}$.
The diagonal reflection matrix of IRS is denoted by $\mathbf{\Theta} = \text{diag} \left\lbrace \lambda_1 e^{j\theta_1}, \lambda_2 e^{j\theta_2}, \dots, \lambda_M e^{j\theta_M} \right\rbrace$, where $\lambda_m \in [0,1]$ and $\theta_m \in [0,2\pi]$ denote the reflection amplitude\footnote{Without loss of generality, we set $\lambda_m=1, \forall m$ to simplify the analysis in the rate-centric communication networks, where the IRS is usually deployed to enhance the amplitude of the reflective signals.}
and phase shift of the $m$-th element equipped on the IRS, respectively.
The total bandwidth $W$ is divided into $K$ subchannels, denoted by $\mathcal{K}=\{1,2,\dots, K\}$, and all subchannels can be reused among BSs to improve the spectrum efficiency.
In an effort to reduce the decoding complexity of SIC procedure at the receiver, we assume that the number of paired NOMA users, simultaneously sharing the available spectrum in each cell, is no more than $A_{\text{max}}$, while $A_{\text{max}} \ge 2$.
{
The investigation of more complicated user grouping schemes over subchannels may further enhance the achievable performance of the considered networks at the cost of complexity, but the performance comparison of different user grouping schemes is beyond the scope of this paper.
Additionally, it is an interesting and meaningful research topic to adopt a proper user grouping scheme under different scenarios and requirements, which should be included in our future work with the aid of the results derived in this paper.}


In the NOMA downlink transmission, let $\alpha_{ij} \in \{0,1\}$ and $\beta_{jk} \in \{0,1\}$ denote the user association indicator and subchannel assignment factor, respectively.
Specifically, we have $\alpha_{ij}=1$ if the $i$-th user is associated with the $j$-th BS, otherwise $\alpha_{ij}=0$.
Furthermore, we have $\beta_{jk}=1$ if the $k$-th subchannel is assigned to the $j$-th BS, otherwise $\beta_{jk}=0$.
Hence, the $i$-th user will be served by the $j$-th BS on the $k$-th subchannel if and only if $\alpha_{ij}\beta_{jk}=1$, otherwise $\alpha_{ij}\beta_{jk}=0$.
Then, the superimposed signal, $x_{jk}$, broadcasted by the $j$-th BS on the $k$-th subchannel can be given by
\begin{equation}\label{superimposed_message}
x_{jk} = \underbrace{\alpha_{ij}\beta_{jk} \sqrt{p_{ijk}} x_{ijk}}_{\mathbf{signal~for~user~i}} \ + \ \underbrace{ \sum \nolimits_{t=1, t \ne i}^{I} \alpha_{tj}\beta_{jk} \sqrt{p_{tjk}} x_{tjk}}_{\mathbf{signal~for~other~paired~users}},
\end{equation}
where $x_{ijk}$ and $p_{ijk}$ denote the signal and power transmitted by BS $j$ on subchannel $k$ for user $i$, respectively.

Considering the intra-cell and inter-cell interference on the $k$-th subchannel \cite{Pan2020Multicell, Zuo2020Intelligent, Xie2020Max}, the received signal of user $i$ associated with BS $j$ on subchannel $k$ is expressed as\footnote{The channels in this work are considered as frequency-flat fading (or sometimes called quasi-static channels)  such as in \cite{Zhao2017Spectrum, Hou2020Reconfigurable, Zhang2020Capacity, Cheng2020Downlink, Zhang2020Reconfigurable}, the effect of multi-path delay is very small and there is no inter-symbol interference. Thus, it is reasonable to ignore the time delay difference between direct and reflective links.}
\begin{eqnarray}\label{received_signal}
y_{ijk}
&=&\underbrace{\left( h_{ijk} + \mathbf{g}_{ik}^{H} \mathbf{\Theta} \mathbf{f}_{jk}\right)  \alpha_{ij}\beta_{jk} \sqrt{p_{ijk}} x_{ijk}}_{\mathbf {desired~signal}}\nonumber
+ \underbrace{\left( h_{ijk} + \mathbf{g}_{ik}^{H} \mathbf{\Theta} \mathbf{f}_{jk}\right)  \sum \nolimits_{t=1, t \ne i}^{I} \alpha_{tj}\beta_{jk} \sqrt{p_{tjk}} x_{tjk}}_{\mathbf {intra-cell~interference}}  \nonumber \\
&+& \underbrace{\sum \nolimits_{s=1, s \ne j}^{J} \left( h_{isk} + \mathbf{g}_{ik}^{H} \mathbf{\Theta} \mathbf{f}_{sk}\right) \sum \nolimits_{t=1}^{I} \alpha_{ts}\beta_{sk} \sqrt{p_{tsk}} x_{tsk}}_{\mathbf {inter-cell~interference}} \ + \  \underbrace{z_{ijk}}_{\mathbf {noise}},
\end{eqnarray}
where
$h_{ijk}$ denotes the Rayleigh fading channel between BS $j$ and user $i$ on subchannel $k$ \cite{Zuo2020Resource},
$\mathbf{f}_{jk} \in \mathbb{C}^{M \times 1}$ represents Rician fading channels between BS $j$ and IRS on subchannel $k$ \cite{Zhang2020Capacity},
$\mathbf{g}_{ik} \in \mathbb{C}^{M \times 1}$ formulates Rayleigh fading channels between IRS and user $i$ on subchannel $k$ \cite{Xie2020Max},
and $z_{ijk}$ indicates the additive white Gaussian noise (AWGN) with zero mean and variance $\sigma^2$, i.e., $z_{ijk} \sim \mathcal{CN}(0,\sigma^2)$.

We denote the SIC decoding order as $\pi_{jk}(i)$ for the user $i$ associated with BS $j$ on subchannel $k$.
Specifically, we have $\pi_{jk}(i)=n$ if the message of user $i$ is the $n$-th signal to be decoded at the receiver, namely, user $i$ first decodes the signals of all the previous $(n-1)$ users, and then successively subtracts their signals to decode its own desired signal. 
For example, two users $i$ and $\tilde{i}$ associated with BS $j$ on subchannel $k$, satisfying $\pi_{jk}(i) \le \pi_{jk}(\tilde{i})$, user $\tilde{i}$ is capable of successfully canceling interference from the superposition signal of user $i$ with the aid of SIC.
Let $H_{ijk}=h_{ijk} + \mathbf{g}_{ik}^{H} \mathbf{\Theta} \mathbf{f}_{jk}$ denote the combined channel gain,
and $P_{ijk}=\alpha_{ij}\beta_{jk} p_{ijk}$ represents the power allocation.
{
Then, the decoding order constraints for guaranteeing success SIC can be formulated as the equation (\ref{SIC_condition_1}).

\begin{equation}\label{SIC_condition_1}
\frac{ | H_{\tilde{i}jk} | ^2 P_{ijk} }{ | H_{\tilde{i}jk} | ^2 \sum \nolimits_{ \pi_{jk}(\hat{i}) > \pi_{jk}(i) } P_{\hat{i}jk} + I_{\tilde{i}jk}^{\text{inter}}+ \sigma ^2 }
\ge
\frac{ | H_{ijk} | ^2 P_{ijk} }{ I_{ijk}^{\text{intra}} + I_{ijk}^{\text{inter}} + \sigma ^2 },
\end{equation}
where
$I_{ijk}^{\text{intra}} = \left| H_{ijk} \right| ^2 \sum_{ \pi_{jk}(\hat{i}) > \pi_{jk}(i) } P_{\hat{i}jk}$
and
$I_{ijk}^{\text{inter}} = \sum_{s=1, s \ne j}^{J} \left| H_{isk} \right| ^2 \sum_{t=1}^{I} P_{tsk}$
are the intra-cell and inter-cell interference, respectively.
}

It indicates that the achievable SINR of user $\tilde{i}$ to decode user $i$ is no less than that of user $i$.
By simple operations, the inequality (\ref{SIC_condition_1}) can be reformulated as the equation (\ref{SIC_condition_2}).

\begin{equation}\label{SIC_condition_2}
\Delta_{jk}(i,\tilde{i}) =
| H_{\tilde{i}jk} | ^2 \left( I_{ijk}^{\text{inter}} + \sigma ^2 \right)
-
| H_{ijk} | ^2 \left( I_{\tilde{i}jk}^{\text{inter}} + \sigma ^2 \right) \ge 0, {\rm \ if \ }  \pi_{jk}(i) \le \pi_{jk}(\tilde{i}).
\end{equation}

%
Accordingly, the received SINR of user $i$ associated with BS $j$ on subchannel $k$ is given by
\begin{eqnarray}\label{SINR}
\text{SINR}_{ijk} =
\frac{ \left| H_{ijk} \right| ^2 P_{ijk} }{ I_{ijk}^{\text{intra}} + I_{ijk}^{\text{inter}} + \sigma ^2 }.
\end{eqnarray}

Therefore, the corresponding achievable downlink data rate of user $i$ associated with BS $j$ on subchannel $k$ is calculated as
\begin{equation}\label{data_rate}
R_{ijk}=\frac{W}{K} \log_{2} \left( 1 + \frac{ \left| H_{ijk} \right| ^2 P_{ijk} }{ I_{ijk}^{\text{intra}} + I_{ijk}^{\text{inter}} + \sigma ^2 } \right).
\end{equation}

\subsection{Problem Formulation}
By jointly designing user association, subchannel assignment, reflection matrix, power allocation, and decoding order in the IRS-aided multi-cell NOMA network, the objective of this paper is to maximize the sum rate of users subject to the SIC decoding constraints, the QoS requirements, and the maximum power constraints, etc.
Hence, the optimization problem can be formulated as
{\allowdisplaybreaks[4]
\begin{subequations}
	\label{max_sum_rate}
	\begin{eqnarray}
	\label{variable}
	&{\rm var}  & \left\lbrace \alpha_{ij}, \beta_{jk}, \mathbf{\Theta}, p_{ijk}, \pi_{jk}(i) \ | \ \forall i,j,k \right\rbrace, \\
	\label{max_sum_rate_objective}
	&\max & \sum \nolimits_{i=1}^{I} \sum \nolimits_{j=1}^{J} \sum \nolimits_{k=1}^{K} R_{ijk}, \\
	\label{max_sum_rate_constraint_success_SIC}
	&{\rm s.t.} & \Delta_{jk}(i,\tilde{i}) \ge 0, \ \text{if} \ \pi_{jk}(i) \le \pi_{jk}(\tilde{i}), \\
	\label{max_sum_rate_constraint_min_rate}
	&{}& \sum \nolimits_{j=1}^{J} \sum \nolimits_{k=1}^{K} R_{ijk} \ge R_{\text{min}}, \ \forall i, \\
	\label{max_sum_rate_constraint_max_power}
	&{}& \sum \nolimits_{i=1}^{I} \sum \nolimits_{k=1}^{K} P_{ijk} \le P_{\text{max}}, \ \forall j, \\
	\label{max_sum_rate_constraint_user_association}
	&{}& \sum \nolimits_{j=1}^{J} \alpha_{ij} = 1, \ \forall i, \\
	\label{max_sum_rate_constraint_max_users}
	&{}& 2 \le \sum \nolimits_{i=1}^{I} \alpha_{ij} \le A_{\text{max}}, \ \forall j, \\
	\label{max_sum_rate_constraint_channel_assignment}
	&{}& \sum \nolimits_{k=1}^{K} \beta_{jk} \ge 1, \ \forall j, \\
	\label{max_sum_rate_constraint_channel_reuse}
	&{}& \sum \nolimits_{j=1}^{J} \beta_{jk} \ge 1, \ \forall k, \\
	\label{max_sum_rate_constraint_binary}
	&{}& \alpha_{ij}, \beta_{jk} \in \{0,1\}, \ \forall i,j,k, \\
	\label{max_sum_rate_constraint_reflection}
	&{}& \theta_m \in [0,2\pi], \ \forall m, \\
	\label{max_sum_rate_constraint_power}
	&{}& p_{ijk} \ge 0, \ \forall i,j,k, \\
	\label{max_sum_rate_constraint_decoding_order}
	&{}& \pi_{jk} \in \Omega_{jk}, \ \forall j,k,
	\end{eqnarray}
\end{subequations}
where the optimization variables are given in (\ref{variable}),
$R_{\text{min}}$ is the minimum data rate required by each user,
$P_{\text{max}}$ is the maximum transmission power provided by each BS,
{
and $\Omega_{jk}$ is the set of all possible SIC decoding orders\footnote{For example, if there are three users in set $\Omega_{jk}$, indexed by $i_{1}$, $i_{2}$ and $i_{3}$, then the permutations can be given as $\Omega_{jk} = \{  (i_{1} , i_{2} , i_{3}), (i_{1} , i_{3} , i_{2}), (i_{2} , i_{1} , i_{3}), (i_{2} , i_{3} , i_{1}), (i_{3} , i_{1} , i_{2}), (i_{3} , i_{2} , i_{1}) \}$.}
for the users associated with the $j$-th cell on the $k$-th subchannel.}
Constraint (\ref{max_sum_rate_constraint_success_SIC}) ensures that the SIC decoding can be conducted successfully at the receiver.
Constraint (\ref{max_sum_rate_constraint_min_rate}) guarantees that the QoS requirement of each user is satisfied.
Constraint (\ref{max_sum_rate_constraint_max_power}) denotes the available transmission power for BSs.
Constraint (\ref{max_sum_rate_constraint_user_association}) represents that each user is associated with one BS.
Constraint (\ref{max_sum_rate_constraint_max_users}) indicates that the number of users multiplexed in each cell is no less than two, and no more than $A_{\text{max}}$.
Constraints (\ref{max_sum_rate_constraint_channel_assignment})-(\ref{max_sum_rate_constraint_channel_reuse}) describe that each BS is assigned with at least one subchannel, and vice versa.
Constraints (\ref{max_sum_rate_constraint_binary})-(\ref{max_sum_rate_constraint_decoding_order}) are invoked for restricting the indication factors, phase shifts, transmit power and decoding order, respectively.}

\begin{figure*}[t]
	\centering
	\includegraphics[width=5 in]{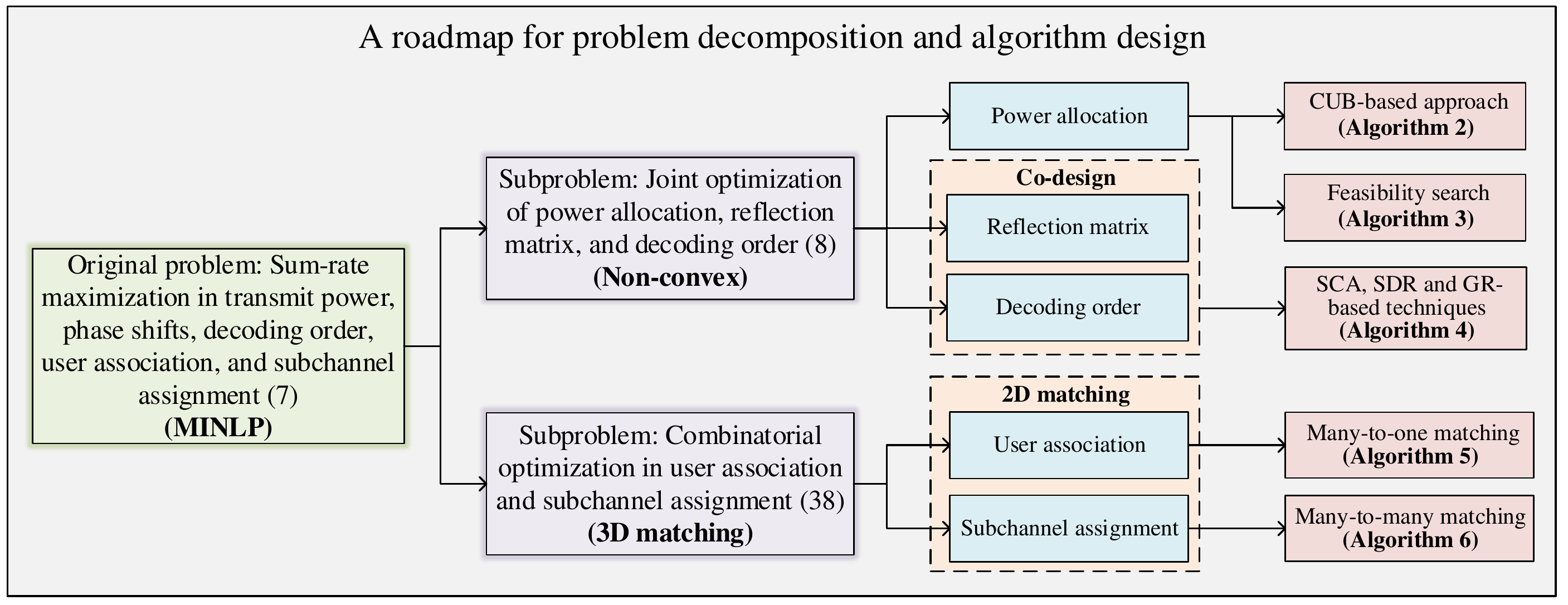}
	\caption{A roadmap for the problem decomposition and the proposed algorithms to subproblems.}
	\label{algorithm_overview}
\end{figure*}

Due to the existence of integer variables $\alpha_{ij}$, $\beta_{jk}$ and the continuous variables $p_{ijk}$, $\mathbf{\Theta}$, as well as their highly coupling at the non-convex objective function and constraints, It can be observed that the sum-rate maximization problem (\ref{max_sum_rate}) is a MINLP problem, which is NP-hard \cite{Cui2018Optimal} and is non-trivial to solve optimally by common standard optimization approaches.
Additionally, the exhaustive search is not feasible, since the computational complexity grows exponentially over the total number of variables.
Therefore, it is essential to transform problem (\ref{max_sum_rate}) into some tractable convex subproblems, which can be solved separately and alternatively over iterations.
{
To this end, the alternating optimization method can be invoked as an intuitive approach to solve the original MINLP problem (\ref{max_sum_rate}) in an efficient manner.

\subsection{Algorithm Overview}
Fig. \ref{algorithm_overview} gives an overview of the roadmap for the proposed problem decomposition and the designed algorithms to the corresponding subproblems.
As one can see, the challenging problem (\ref{max_sum_rate}) is decomposed into a non-convex problem (\ref{problem_power_reflection_order}) and a combinatorial optimization problem (\ref{3D_matching}), which are efficiently solved by the optimization methods and matching theory in Section \ref{optimization} and Section \ref{matching}, respectively.

For an overview, the designed alternating optimization algorithm for sum-rate maximization in IRS-aided multi-cell NOMA networks is summarized in \textbf{Algorithm \ref{alternating_algorithm}}.
Specifically, the initial points are found by the feasibility-searching algorithm, i.e., \textbf{Algorithm \ref{algorithm_feasibility_searching}}.
In the first step, power allocation is performed based on \textbf{Algorithm \ref{algorithm_1}}.
In the second step, the co-design of reflection matrix and decoding order is obtained according to \textbf{Algorithm \ref{codesign_of_reflection_decoding}}.
In the third step, the user association is executed by \textbf{Algorithm \ref{algorithm_4}}.
In the fourth step, the subchannel assignment is conducted via \textbf{Algorithm \ref{algorithm_5}}.
Then, the above steps are performed alternatively until converge.
Accordingly, the complexity and convergence of the four-step \textbf{Algorithm \ref{alternating_algorithm}} are provided in the following remark.

%
%

\begin{algorithm}[t]
	\caption{Alternating Optimization Algorithm for IRS-Aided Multi-Cell Networks}
	\label{alternating_algorithm}
	\begin{algorithmic}[1]
		{
		\renewcommand{\algorithmicrequire}{\textbf{Initialize}}
		\renewcommand{\algorithmicensure}{\textbf{Output}}
		\STATE \textbf{Initialize} a feasible solution $( \boldsymbol{\alpha}^{(0)}, \boldsymbol{\beta}^{(0)} )$, the tolerance $\epsilon$, the maximum iteration number $N_{1}$, and set the current iteration number as $n_{1}=0$.
		\STATE Randomly generate $( \mathbf{p}^{(0)}, \boldsymbol{\gamma}^{(0)}, \boldsymbol{\Theta}^{(0)}, \boldsymbol{\pi}^{(0)} )$, update the initial points $(\mathbf{p}^{(0)}, \boldsymbol{\gamma}^{(0)})$ by solving problem (\ref{problem_power_lambda_feasibility}) via \textbf{Algorithm \ref{algorithm_feasibility_searching}};
		\REPEAT
		\STATE \textbf{Step 1:} Power allocation 
		\STATE With given $( \boldsymbol{\alpha}^{(n_{1})}, \boldsymbol{\beta}^{(n_{1})}, \mathbf{\Theta}^{(n_{1})}, \boldsymbol{\pi}^{(n_{1})} )$, update $(\mathbf{p}^{(n_{1}+1)}, \boldsymbol{\gamma}^{(n_{1}+1)})$ via \textbf{Algorithm \ref{algorithm_1}};
		\STATE \textbf{Step 2:} Co-design of reflection matrix and decoding order
		\STATE With given $( \boldsymbol{\alpha}^{(n_{1})}, \boldsymbol{\beta}^{(n_{1})}, \mathbf{p}^{(n_{1}+1)}, \boldsymbol{\gamma}^{(n_{1}+1)})$, update $( \mathbf{\Theta}^{(n_{1}+1)}, \boldsymbol{\pi}^{(n_{1}+1)} )$ via \textbf{Algorithm \ref{codesign_of_reflection_decoding}};
		\STATE \textbf{Step 3:} User association
		\STATE With given $( \boldsymbol{\beta}^{(n_{1})}, \mathbf{p}^{(n_{1}+1)}, \mathbf{\Theta}^{(n_{1}+1)}, \boldsymbol{\pi}^{(n_{1}+1)} )$, obtain $\boldsymbol{\alpha}^{(n_{1}+1)}$ via \textbf{Algorithm \ref{algorithm_4}};
		\STATE \textbf{Step 4:} Subchannel assignment
		\STATE With given $( \boldsymbol{\alpha}^{(n_{1}+1)}, \mathbf{p}^{(n_{1}+1)}, \mathbf{\Theta}^{(n_{1}+1)}, \boldsymbol{\pi}^{(n_{1}+1)})$, obtain $\boldsymbol{\beta}^{(n_{1}+1)}$ via \textbf{Algorithm \ref{algorithm_5}}.	
		\STATE Update $n_{1} := n_{1} + 1$;
		\UNTIL the objective value of problem (\ref{max_sum_rate}) converges or $n_{1} > N_{1}$;
		\STATE \textbf{Output} the converged solution $( \boldsymbol{\alpha}^{(n_{1})}, \boldsymbol{\beta}^{(n_{1})}, \mathbf{p}^{(n_{1})}, \mathbf{\Theta}^{(n_{1})}, \boldsymbol{\pi}^{(n_{1})})$.}
	\end{algorithmic}
\end{algorithm}

\begin{remark} \label{complexity_and_convergence}
	\emph{The computational complexity of \textbf{Algorithm \ref{alternating_algorithm}} can be given as
		$\mathcal{O} ( N_3 (2IJK)^3 + N_1 N_2 (2IJK)^3 + N_1 ( (M+4IJK)^6 + N_4 T_{GR} ) + N_1 (IJ^2 + A_{\text{max}}IJN_{it}) + N_1 J^2 K^2(\bar{N}_{it}+1) )$.
		Meanwhile, the proposed \textbf{Algorithm \ref{alternating_algorithm}} is guaranteed to converge as long as the the maximum iteration number $N_{1}$ is set sufficiently large.}
\end{remark}

\begin{IEEEproof}
	Please refer to Appendix \ref{proof_of_remark_1}.
\end{IEEEproof}
}


\section{Joint Optimization of Power, Reflection, and Decoding Order}
\label{optimization}
Given user association and subchannel assignment, we first aim to solve the joint optimization problem of power allocation, reflection matrix, and decoding order, which can be expressed as
\begin{subequations}
	\label{problem_power_reflection_order}
	\begin{eqnarray}
	\label{objective_power_reflection_order}
	&\max \limits_{\mathbf{p},\mathbf{\Theta},\boldsymbol{\pi}} & \sum \nolimits_{i} \sum \nolimits_{j} \sum \nolimits_{k} R_{ijk}, \\
	&{\rm s.t.} & {\rm (\ref{max_sum_rate_constraint_success_SIC}), (\ref{max_sum_rate_constraint_min_rate}), (\ref{max_sum_rate_constraint_max_power}), (\ref{max_sum_rate_constraint_reflection}), (\ref{max_sum_rate_constraint_power}), (\ref{max_sum_rate_constraint_decoding_order})},
	\end{eqnarray}
\end{subequations}
where $\mathbf{p}=\left\lbrace p_{ijk} | \forall i,j,k \right\rbrace $ is the power allocation profile, and $\boldsymbol{\pi}=\left\lbrace \pi_{jk} | \forall j,k \right\rbrace $ is the profile of SIC decoding order.
Due to the intra-cell interference $I_{ijk}^{\text{intra}}$ and the inter-cell interference $I_{ijk}^{\text{inter}}$ in both the objective function (\ref{objective_power_reflection_order}) and the constraints (\ref{max_sum_rate_constraint_success_SIC})-(\ref{max_sum_rate_constraint_min_rate}), it is intractable to solve this non-linear and non-convex problem (\ref{problem_power_reflection_order}) by using the standard convex optimization approaches.

{
In order to address the non-concavity of $R_{ijk}$, we introduce an auxiliary variable set $\boldsymbol{\gamma} = \left\lbrace \gamma_{ijk} | \text{SINR}_{ijk} = \gamma_{ijk}, \forall i,j,k \right\rbrace $, and then the problem (\ref{problem_power_reflection_order}) can be reformulated as
\begin{subequations}
	\label{problem_power_reflection_order_lambda}
	\begin{eqnarray}
	\label{problem_power_reflection_order_lambda_objective}
	&\max \limits_{\mathbf{p},\boldsymbol{\gamma},\mathbf{\Theta},\boldsymbol{\pi}} & \sum \nolimits_{i} \sum \nolimits_{j} \sum \nolimits_{k} \frac{W}{K} \log_{2} \left( 1 + \gamma_{ijk} \right), \\
	\label{problem_power_reflection_order_lambda_constraint_min_rate}
	&{\rm s.t.} & \sum \nolimits_{j} \sum \nolimits_{k} \frac{W}{K} \log_{2} \left( 1 + \gamma_{ijk} \right) \ge R_{\text{min}}, \ \forall i, \\
	\label{problem_power_reflection_order_lambda_constraint_SINR}
	&{}& \text{SINR}_{ijk} \ge \gamma_{ijk}, \ \forall i,j,k, \\
	&{}& {\rm (\ref{max_sum_rate_constraint_success_SIC}), (\ref{max_sum_rate_constraint_max_power}), (\ref{max_sum_rate_constraint_reflection}), (\ref{max_sum_rate_constraint_power}), (\ref{max_sum_rate_constraint_decoding_order})}.
	\end{eqnarray}
\end{subequations}

\begin{remark} \label{remark_equivalent}
	\emph{The optimal solution of problem (\ref{problem_power_reflection_order_lambda}) is also optimal for problem (\ref{problem_power_reflection_order}), due to the fact that problem (\ref{problem_power_reflection_order_lambda}) is equivalent to problem (\ref{problem_power_reflection_order}).}
\end{remark}

\begin{IEEEproof}
	According to Appendix D in \cite{Zuo2020Resource}, the proof of Remark \ref{remark_equivalent} can be derived similarly, which is thus omitted here for brevity.
\end{IEEEproof}

\subsection{Power Allocation}
Given the reflection matrix and decoding order in problem (\ref{problem_power_reflection_order_lambda}), the power allocation subproblem can be given by
\begin{subequations}
	\label{problem_power_lambda}
	\begin{eqnarray}
	&\max \limits_{\mathbf{p},\boldsymbol{\gamma}} & \sum \nolimits_{i} \sum \nolimits_{j} \sum \nolimits_{k} \frac{W}{K} \log_{2} \left( 1 + \gamma_{ijk} \right), \\
	&{\rm s.t.}& {\rm (\ref{max_sum_rate_constraint_success_SIC}),
		(\ref{max_sum_rate_constraint_max_power}),
		(\ref{max_sum_rate_constraint_power}),
		(\ref{problem_power_reflection_order_lambda_constraint_min_rate}),
		(\ref{problem_power_reflection_order_lambda_constraint_SINR})}. 
	\end{eqnarray}
\end{subequations}

Note that the constraint (\ref{max_sum_rate_constraint_success_SIC}) can be equivalently expressed as
\begin{equation} \label{problem_power_constraint_success_SIC}
\sum \limits_{s \ne j}^{J} \left( \left| H_{\tilde{i}jk} \right| ^2 \left| H_{isk} \right| ^2 - \left| H_{ijk} \right| ^2 \left| H_{\tilde{i}sk} \right| ^2 \right) \sum\limits_{t = 1}^I P_{tsk}
+
\left( \left| H_{\tilde{i}jk} \right| ^2 - \left| H_{ijk} \right| ^2 \right) \sigma ^2
\ge
0, {\rm \ if \ }  \pi_{jk}(i) \le \pi_{jk}(\tilde{i}).
\end{equation}

Since the constraint in (\ref{problem_power_constraint_success_SIC}) is linear w.r.t. power allocation $\mathbf{p}$, it is convex for problem (\ref{problem_power_lambda}) with a given user association and subchannel assignment.
At this point, it can be noticed that all constraints in problem (\ref{problem_power_lambda}) are convex excluding the constraint (\ref{problem_power_reflection_order_lambda_constraint_SINR}), which is rewritten as
\begin{equation}
\label{problem_power_lambda_constraint_SINR}
P_{ijk} \ge \gamma_{ijk} \hat{P}_{ijk} + \gamma_{ijk} \bar{P}_{ijk} + \gamma_{ijk} \xi_{ijk},
\end{equation}
where
$\hat{P}_{ijk} = \sum_{ \pi_{jk}(\hat{i}) > \pi_{jk}(i) } P_{\hat{i}jk}$,
$\bar{P}_{ijk} = \sum \limits_{s \ne j} \frac{ \left| H_{isk} \right| ^2}{ \left| H_{ijk} \right| ^2} \sum\nolimits_{t= 1}^I P_{tsk}$,
and
$\xi_{ijk} = \frac{\sigma ^2}{|H_{ijk}|^2}$.
It is worth noting that the product terms $\gamma_{ijk} \hat{P}_{ijk}$ and $\gamma_{ijk} \bar{P}_{ijk}$ are both non-convex, and thus the SINR constraint in (\ref{problem_power_lambda_constraint_SINR}) is still not a convex constraint.
Therefore, it is necessary to transform the constraint (\ref{problem_power_lambda_constraint_SINR}) into a convex one.

Let $f(\gamma_{ijk},\hat{P}_{ijk}) = \gamma_{ijk} \hat{P}_{ijk}$, while $\gamma_{ijk}, \hat{P}_{ijk} \ge 0$.
By replacing $f(\gamma_{ijk},\hat{P}_{ijk})$ with its convex upper bound (CUB) \cite{Tran2012Fast}, the resulting constraint becomes convex\footnote{{According to the proof of Lemma 1 in \cite{Xie2020Joint}, the tightness of CUB can be derived similarly, which is thus omitted here for brevity.}}.
To this end, we define the following function
\begin{equation} \label{problem_power_lambda_constraint_SINR_CUB_1}
g(\gamma_{ijk},\hat{P}_{ijk},\lambda_{ijk}) = \frac{\lambda_{ijk}}{2} \gamma_{ijk}^2 + \frac{1}{2\lambda_{ijk}} \hat{P}_{ijk}^2,
\end{equation}
where $\boldsymbol{\lambda} = \{\lambda_{ijk} | \forall i,j,k\}$ is a coefficient set.
It can be proved that (\ref{problem_power_lambda_constraint_SINR_CUB_1}) is a convex function, 
and $g(\gamma_{ijk},\hat{P}_{ijk},\lambda_{ijk}) \ge f(\gamma_{ijk},\hat{P}_{ijk})$ is satisfied for all $\lambda_{ijk}>0$ \cite{Zuo2020Resource, Tran2012Fast}.
Moreover, it can be derived that the equation will turn to equality when $\lambda_{ijk} = {\hat{P}_{ijk}}/{\gamma_{ijk}}$.
Similarly, let $\bar{f} (\gamma_{ijk},\bar{P}_{ijk}) = \gamma_{ijk} \bar{P}_{ijk}$, then its corresponding CUB can be given as
\begin{equation} \label{problem_power_lambda_constraint_SINR_CUB_2}
\bar{g}(\gamma_{ijk},\bar{P}_{ijk},\bar{\lambda}_{ijk}) = \frac{\bar{\lambda}_{ijk}}{2} \gamma_{ijk}^2 + \frac{1}{2 \bar{\lambda}_{ijk}} \bar{P}_{ijk}^2,
\end{equation}
where $\boldsymbol{\bar{\lambda}} = \{\bar{\lambda}_{ijk} | \forall i,j,k\}$ is the coefficient set for $\bar{g}$.
Therefore, by replacing $f$ and $\bar{f}$ with (\ref{problem_power_lambda_constraint_SINR_CUB_1}) and (\ref{problem_power_lambda_constraint_SINR_CUB_2}), constraint (\ref{problem_power_lambda_constraint_SINR}) can be transformed into the following convex one.
\begin{equation}
\label{problem_power_lambda_constraint_SINR_convex}
P_{ijk} \ge
\frac{1}{2} \left( \lambda_{ijk} + \bar{\lambda}_{ijk} \right) \gamma_{ijk}^2 + \frac{1}{2\lambda_{ijk}} \hat{P}_{ijk}^2 + \frac{1}{2 \bar{\lambda}_{ijk}} \bar{P}_{ijk}^2 + \gamma_{ijk} \xi_{ijk}.
\end{equation}

Next, by replacing (\ref{problem_power_reflection_order_lambda_constraint_SINR}) with its approximate constraint (\ref{problem_power_lambda_constraint_SINR_convex}), it can be observed that both the objective function and all constraints in problem (\ref{problem_power_lambda}) become convex, and hence the Karush-Kuhn-Tucker (KKT) solution of (\ref{problem_power_lambda}) can be iteratively updated until convergence by optimally solving its convex approximation problem with CVX.
The details of the proposed CUB-based power allocation algorithm with an adjustable convergence accuracy $\epsilon$ are summarized in \textbf{Algorithm \ref{algorithm_1}}, where the fixed coefficients $\gamma_{ijk}$ and $\bar{\gamma}_{ijk}$ in the $n_{2}$-th iteration can be updated by
\begin{eqnarray}
\label{CUB_update_lambda_1}
\lambda_{ijk}^{(n_{2})} := {\hat{P}_{ijk}^{(n_{2}-1)}}/{\gamma_{ijk}^{(n_{2}-1)}}, \\
\label{CUB_update_lambda_2}
\bar{\lambda}_{ijk}^{(n_{2})} := {\bar{P}_{ijk}^{(n_{2}-1)}}/{\gamma_{ijk}^{(n_{2}-1)}}.
\end{eqnarray}

\begin{algorithm}[h]
	\caption{CUB-Based Algorithm for Power Allocation}
	\label{algorithm_1}
	\begin{algorithmic}[1]
		\renewcommand{\algorithmicrequire}{\textbf{Initialize}}
		\renewcommand{\algorithmicensure}{\textbf{Output}}
		\STATE \textbf{Initialize} $\mathbf{p}^{(0)}$ and $\boldsymbol{\gamma}^{(0)}$, the tolerance $\epsilon$, the maximum iteration number $N_{2}$, and set the current iteration number as $n_{2}=0$.
		\STATE Compute utility $U^{(0)} = \sum_{i} \sum_{j} \sum_{k} \frac{W}{K} \log_{2} ( 1 + \gamma_{ijk}^{(0)} )$;
		\REPEAT
		\STATE With given $\mathbf{p}^{(n_{2})}$ and $\boldsymbol{\gamma}^{(n_{2})}$, update $\boldsymbol{\lambda}^{(n_{2}+1)}$ and $\boldsymbol{\bar{\lambda}}^{(n_{2}+1)}$ by using (\ref{CUB_update_lambda_1}) and (\ref{CUB_update_lambda_2});
		\STATE With obtained $\boldsymbol{\lambda}^{(n_{2}+1)}$ and $\boldsymbol{\bar{\lambda}}^{(n_{2}+1)}$, compute $\mathbf{p}^{(n_{2}+1)}$ and $\boldsymbol{\gamma}^{(n_{2}+1)}$ by solving the substituted problem of (\ref{problem_power_lambda}) with CVX;
		\STATE With obtained $\boldsymbol{\gamma}^{(n_{2}+1)}$, calculate $U^{(n_{2}+1)}=U( \boldsymbol{\gamma}^{(n_{2}+1)} ) $;
		\STATE Update $n_{2} := n_{2} + 1$;
		\UNTIL $|U^{(n_{2})}-U^{(n_{2}-1)}| < \epsilon $ or $n_{2} > N_{2}$;
		\STATE \textbf{Output} the converged solutions $\mathbf{p}^*$ and $\boldsymbol{\gamma}^{*}$.
	\end{algorithmic}
\end{algorithm}

Usually, it is non-trivial and difficult to find the  initial feasible points $\mathbf{p}^{(0)}$ and $\boldsymbol{\gamma}^{(0)}$ in \textbf{Algorithm \ref{algorithm_1}}.
To reduce the sensitivity of \textbf{Algorithm \ref{algorithm_1}} to the feasibility of the initial points, we formulate a new error minimization problem and propose a novel feasibility-searching algorithm.
Let $\bar{\epsilon} \ge 0$ denotes the distance (also called `feasibility error') from the initial points to the feasible domain, then the feasible point search problem can be developed as \cite{Zuo2020Resource}
\begin{subequations}
	\label{problem_power_lambda_feasibility}
	\begin{eqnarray}
	&\min \limits_{\mathbf{p},\boldsymbol{\gamma}, \bar{\epsilon}} & \bar{\epsilon}, \\
	&{\rm s.t.}& \sum \nolimits_{s \ne j} H \sum \nolimits_{t = 1}^I P_{tsk} +	\left( \left| H_{\tilde{i}jk} \right| ^2 - \left| H_{ijk} \right| ^2 \right) \sigma ^2 + \bar{\epsilon} \ge 0, \\
	&{}& \sum \nolimits_{j} \sum \nolimits_{k} \frac{W}{K} \log_{2} \left( 1 + \gamma_{ijk} \right) + \bar{\epsilon} \ge R_{\text{min}}, \ \forall i, \\
	&{}& P_{ijk} + \bar{\epsilon} \ge \frac{1}{2} \left( \lambda_{ijk} + \bar{\lambda}_{ijk} \right) \gamma_{ijk}^2 + \frac{1}{2\lambda_{ijk}} \hat{P}_{ijk}^2 + \frac{1}{2 \bar{\lambda}_{ijk}} \bar{P}_{ijk}^2 + \gamma_{ijk} \xi_{ijk}, \\
	&{}& \sum \nolimits_{i=1}^{I} \sum \nolimits_{k=1}^{K} P_{ijk} \le P_{\text{max}} + \bar{\epsilon}, \forall j, {\rm \ and \ } \bar{\epsilon} \ge 0.
	\end{eqnarray}
\end{subequations}
where $H = \left| H_{\tilde{i}jk} \right| ^2 \left| H_{isk} \right| ^2 - \left| H_{ijk} \right| ^2 \left| H_{\tilde{i}sk} \right| ^2$.
Note that the substituted problem of (\ref{problem_power_lambda}) and problem (\ref{problem_power_lambda_feasibility}) have the same set of feasible $\mathbf{p}$ and $\boldsymbol{\gamma}$, while problem (\ref{problem_power_lambda_feasibility}) is more robust than the former in terms of the initial solution, e.g., it does not request initial points $\mathbf{p}^{(0)}$ and $\boldsymbol{\gamma}^{(0)}$ in the feasible domain.
Furthermore, it is worth pointing out that problem (\ref{problem_power_lambda_feasibility}) is a jointly convex optimization problem, which can be efficiently solved by CVX as well.
As a result, the proposed feasibility-searching method is given in \textbf{Algorithm \ref{algorithm_feasibility_searching}}.

\begin{algorithm}[h]
	\caption{Feasibility-Searching Algorithm for Solving Problem (\ref{problem_power_lambda_feasibility})}
	\label{algorithm_feasibility_searching}
	\begin{algorithmic}[1]
		\renewcommand{\algorithmicrequire}{\textbf{Initialize}}
		\renewcommand{\algorithmicensure}{\textbf{Output}}
		\STATE \textbf{Initialize} $\mathbf{p}^{(0)}$ and $\boldsymbol{\gamma}^{(0)}$ randomly, the tolerance $\epsilon$, the maximum iteration number $N_{3}$, and set the current iteration number as $n_{3}=0$.
		\REPEAT
		\STATE With given $\mathbf{p}^{(n_{3})}$ and $\boldsymbol{\gamma}^{(n_{3})}$, update $\boldsymbol{\lambda}^{(n_{3}+1)}$ and $\boldsymbol{\bar{\lambda}}^{(n_{3}+1)}$ by using (\ref{CUB_update_lambda_1}) and (\ref{CUB_update_lambda_2});
		\STATE With updated $\boldsymbol{\lambda}^{(n_{3}+1)}$ and $\boldsymbol{\bar{\lambda}}^{(n_{3}+1)}$, compute $\mathbf{p}^{(n_{3}+1)}$, $\boldsymbol{\gamma}^{(n_{3}+1)}$ and $\bar{\epsilon}$ by solving (\ref{problem_power_lambda_feasibility});
		\STATE Update $n_{3} := n_{3} + 1$;
		\UNTIL $\bar{\epsilon}$ below the tolerance $\epsilon $ or $n_{3} > N_{3}$;
		\STATE \textbf{Output} the obtained solution $\mathbf{p}^{(n_{3})}$ and $\boldsymbol{\gamma}^{(n_{3})}$;
	\end{algorithmic}
\end{algorithm}

\begin{remark}\emph{Unlike \textbf{Algorithm \ref{algorithm_1}}, the initial points $\mathbf{p}^{(0)}$ and $\boldsymbol{\gamma}^{(0)}$ in \textbf{Algorithm \ref{algorithm_feasibility_searching}} can be generated randomly.
		When $\bar{\epsilon} = 0$, the optimal solutions of (\ref{problem_power_lambda_feasibility}) are feasible for the substituted problem of (\ref{problem_power_lambda}).
		Hence, the output of \textbf{Algorithm \ref{algorithm_feasibility_searching}} can be used as the initial input of \textbf{Algorithm \ref{algorithm_1}}.}
\end{remark}

\subsection{Co-design of Reflection Matrix and Decoding Order}
With the converged results $\mathbf{p}^*$ and $\boldsymbol{\gamma}^{*}$ obtained from \textbf{Algorithm \ref{algorithm_1}}, problem (\ref{problem_power_reflection_order_lambda}) is simplified into the following feasibility-check subproblem
\begin{subequations}
	\label{problem_feasibility_check_theta_pi}
	\begin{eqnarray}
	&\text{find} & \mathbf{\Theta}, \boldsymbol{\pi} \\
	&{\rm s.t.} & {\rm (\ref{max_sum_rate_constraint_success_SIC}),
		(\ref{max_sum_rate_constraint_reflection}),
		(\ref{max_sum_rate_constraint_decoding_order}),
		(\ref{problem_power_reflection_order_lambda_constraint_SINR})}.
	\end{eqnarray}
\end{subequations}

Due to the coupling of $\mathbf{\Theta}$ and $\boldsymbol{\pi}$ in constraint (\ref{max_sum_rate_constraint_success_SIC}) and the non-convexity of constraint (\ref{problem_power_reflection_order_lambda_constraint_SINR}), problem (\ref{problem_feasibility_check_theta_pi}) is non-convex and difficult to be solved directly.
Alternatively, in order to make this problem tractable, we first consider the problem of reflection matrix design under a determined SIC decoding order, then in turn a low-complexity algorithm for decoding order determination is designed based on the combined channel gain tuned by the IRS.

Specifically, with given $\boldsymbol{\pi}$, the problem (\ref{problem_feasibility_check_theta_pi}) is reduced to
\begin{subequations}
	\label{problem_feasibility_check_theta}
	\begin{eqnarray}
	&\text{find}& \mathbf{\Theta} \\
	&{\rm s.t.}& {\rm (\ref{max_sum_rate_constraint_success_SIC}), (\ref{max_sum_rate_constraint_reflection}), (\ref{problem_power_reflection_order_lambda_constraint_SINR}).}
	\end{eqnarray}
\end{subequations}
where the non-convex constraints in (\ref{max_sum_rate_constraint_success_SIC}) and (\ref{problem_power_reflection_order_lambda_constraint_SINR}) can be equivalently expressed as the following (\ref{problem_feasibility_check_theta_constraint_SIC}) and (\ref{problem_feasibility_check_theta_constraint_SINR}), respectively.
\begin{equation} \label{problem_feasibility_check_theta_constraint_SIC}
\left| H_{\tilde{i}jk} \right|^2 \left( \sum \nolimits_{s \ne j} \left| H_{isk} \right|^2 \sum \nolimits_{t} P_{tsk} + \sigma ^2 \right)
\ge
\left| H_{ijk} \right|^2 \left( \sum \nolimits_{s \ne j} \left| H_{\tilde isk} \right| ^2 \sum \nolimits_{t} P_{tsk} + \sigma ^2 \right),
\end{equation}
\begin{equation} \label{problem_feasibility_check_theta_constraint_SINR}
\left| H_{ijk} \right|^2 \left( P_{ijk} - \gamma _{ijk} \hat{P}_{ijk} \right)
\ge
\sum \nolimits_{s \ne j} \left| H_{isk} \right|^2 \gamma _{ijk} \sum \nolimits_{t} P_{tsk} + \gamma _{ijk} \sigma^2, \ \forall i,j,k.
\end{equation}

First of all, to handle the non-convexity of constraint (\ref{problem_feasibility_check_theta_constraint_SIC}), we introduce the new variables
$a_{\tilde{i}jk} = \left| H_{\tilde{i}jk} \right| ^2$,
$b_{isk} = \left| H_{isk} \right| ^2$,
$c_{ijk} = \left| H_{ijk} \right| ^2$,
and
$d_{\tilde{i}sk} = \left| H_{\tilde{i}sk} \right| ^2$,
then constraint (\ref{problem_feasibility_check_theta_constraint_SIC}) can be reformulated as
\begin{equation} \label{problem_feasibility_check_theta_constraint_SIC_with_abcd}
\sum \nolimits_{s \ne j} a_{\tilde{i}jk} b_{isk} \sum \nolimits_{t} P_{tsk} + a_{\tilde{i}jk} \sigma ^2
\ge
\sum \nolimits_{s \ne j} c_{ijk} d_{\tilde{i}sk} \sum \nolimits_{t} P_{tsk} + c_{ijk} \sigma ^2.
\end{equation}

Then, by adopting an approximation of the difference of two convex functions (DC), the constraint (\ref{problem_feasibility_check_theta_constraint_SIC_with_abcd}) can be approximated by
\begin{align} \label{problem_feasibility_check_theta_constraint_SIC_with_abcd_SCA}
\sum \nolimits_{s \ne j} \left( a_{\tilde{i}jk} b_{isk}^{(n)} + a_{\tilde{i}jk}^{(n)} b_{isk} - a_{\tilde{i}jk}^{(n)} b_{isk}^{(n)} \right) & \sum \nolimits_{t} P_{tsk} + a_{\tilde{i}jk} \sigma ^2 \\
\ge \sum \nolimits_{s \ne j} & \left( c_{ijk} d_{\tilde{i}sk}^{(n)} + c_{ijk}^{(n)} d_{\tilde{i}sk} - c_{ijk}^{(n)} d_{\tilde{i}sk}^{(n)} \right) \sum \nolimits_{t} P_{tsk} + c_{ijk} \sigma ^2, \nonumber
\end{align}
where the left-hand side is the first-order Taylor expansion of $a_{\tilde{i}jk} b_{isk}$ at the point $(a_{\tilde{i}jk}^{(n)}, b_{isk}^{(n)})$ obtained after the $n$-th iteration, and the right-hand side is the first-order Taylor expansion of the non-convex term $c_{ijk} d_{\tilde{i}sk}$ at the point $(c_{ijk}^{(n)}, d_{\tilde{i}sk}^{(n)})$, similarly.

Next, to handle the non-convexity of constraint (\ref{problem_feasibility_check_theta_constraint_SINR}), we first define $\boldsymbol{\rho}_{ijk} = \text{diag}\{ \mathbf{g}_{ik}^{H} \} \mathbf{f}_{jk}$ and $\boldsymbol{\nu} = [\nu_1, \nu_2, \dots, \nu_M]^H$, where $\nu_m = e^{j\theta_m}$.
Then, we have
\begin{equation}
|H_{ijk}|^2
= |h_{ijk} + \mathbf{g}_{ik}^{H} \mathbf{\Theta} \mathbf{f}_{jk}| ^2
= |h_{ijk} + \boldsymbol{\nu}^H \boldsymbol{\rho}_{ijk} | ^2
= \boldsymbol{\bar{\nu}}^H \mathbf{C}_{ijk} \boldsymbol{\bar{\nu}} + |h_{ijk}|^2,
\end{equation}
where
\begin{equation}
\mathbf{C}_{ijk} = \left[ \begin{array}{cc}
\boldsymbol{\rho}_{ijk} \boldsymbol{\rho}_{ijk}^{H} & h_{ijk} \boldsymbol{\rho}_{ijk} \\
h_{ijk} \boldsymbol{\rho}_{ijk}^{H} & 0 
\end{array} 
\right ]
\text{and} \
\boldsymbol{\bar{\nu}} = \left[ \begin{array}{c}
\boldsymbol{{\nu}} \\
1
\end{array} 
\right ].
\end{equation}

Meanwhile, we define $\textbf{V} = \boldsymbol{\bar{\nu}} \boldsymbol{\bar{\nu}}^{H}$, while $\textbf{V} \succeq \textbf{0}$ and $\text{rank}(\textbf{V}) = 1$.
So we have $\boldsymbol{\bar{\nu}}^H \mathbf{C}_{ijk} \boldsymbol{\bar{\nu}} = \text{tr}(\mathbf{C}_{ijk} \textbf{V})$, and then the constraint (\ref{problem_feasibility_check_theta_constraint_SINR}) can be rewritten as the following convex one:
\begin{equation} \label{problem_feasibility_check_theta_constraint_SINR_convex}
\left( \text{tr}\left( \mathbf{C}_{ijk} \textbf{V} \right) + \left| h_{ijk} \right|^2 \right) \tilde{P}_{ijk}
\ge
\sum \nolimits_{s \ne j} \left( \text{tr} \left( \mathbf{C}_{isk} \textbf{V} \right) + \left| h_{isk} \right|^2 \right) \gamma_{ijk} \sum \nolimits_{t} P_{tsk} + \gamma _{ijk} \sigma^2, \ \forall i,j,k,
\end{equation}
where $\tilde{P}_{ijk} = P_{ijk} - \gamma _{ijk} \hat{P}_{ijk}$.

Based on the above approximations, the non-convex problem in (\ref{problem_feasibility_check_theta}) can be reformulated into the following approximated problem:
{\allowdisplaybreaks[4]
\begin{subequations}
	\label{problem_feasibility_check_theta_abcd}
	\begin{eqnarray}
	&\text{find}& \mathbf{V}, \boldsymbol{a}, \boldsymbol{b}, \boldsymbol{c}, \boldsymbol{d} \\
	&{\rm s.t.}& \text{tr}\left( \mathbf{C}_{\tilde{i}jk} \textbf{V} \right) + \left| h_{\tilde{i}jk} \right|^2 \ge a_{\tilde{i}jk} \ge 0, \ \forall \tilde{i},j,k, \\
	&{}&  \text{tr}\left( \mathbf{C}_{isk} \textbf{V} \right) + \left| h_{isk} \right|^2 \ge b_{isk} \ge 0, \ \forall i,s,k \\
	&{}& \text{tr}\left( \mathbf{C}_{ijk} \textbf{V} \right) + \left| h_{ijk} \right|^2 \le c_{ijk}, \ \forall i,j,k \\
	&{}& \text{tr}\left( \mathbf{C}_{\tilde{i}sk} \textbf{V} \right) + \left| h_{\tilde{i}sk} \right|^2 \le d_{\tilde{i}sk}, \ \forall \tilde{i},j,k \\
	&{}& \textbf{V}_{m,m} = 1, \forall m = 1,2,\ldots,M+1, \\
	&{}& \textbf{V} \succeq \textbf{0}, \quad \text{rank}(\textbf{V}) = 1, \quad {\rm (\ref{problem_feasibility_check_theta_constraint_SIC_with_abcd_SCA}) \ and \ (\ref{problem_feasibility_check_theta_constraint_SINR_convex})},
	\end{eqnarray}
\end{subequations}
where
$\boldsymbol{a} = \{ a_{\tilde{i}jk} |  \forall \tilde{i} \in \mathcal{I}, j \in \mathcal{J}, k \in \mathcal{K} \}$,
$\boldsymbol{b} = \{ b_{isk} |  \forall i \in \mathcal{I}, s \ne j, k \in \mathcal{K} \}$,
$\boldsymbol{c} = \{ c_{ijk} |  \forall i \in \mathcal{I}, j \in \mathcal{J}, k \in \mathcal{K} \}$,
and
$\boldsymbol{d} = \{ d_{\tilde{i}sk} |  \forall \tilde{i} \in \mathcal{I}, s \ne j, k \in \mathcal{K} \}$ are the introduced sets of auxiliary variables.}

}



Although the rank-one constraint is still non-convex, the semidefinite relaxation (SDR) can be applied to relax problem (\ref{problem_feasibility_check_theta_abcd}) into a standard semidefinite programming (SDP) problem, and thus the optimal $\mathbf{V}^{*}$ can be obtained by solving the relaxed convex problem with the SeDuMi solver in CVX.
Finally, with $\mathbf{V}^{*} = \boldsymbol{\bar{\nu}}^{*} \boldsymbol{\bar{\nu}}^{*H}$, the optimal reflection matrix $\mathbf{\Theta}^{*}$ is obtained.
However, if $\text{rank}(\textbf{V}) \ne 1$, the Gaussian randomization (GR) method has to be invoked to construct a rank-one solution based on the higher-rank solution of the relaxed problem.

Namely, if $\text{rank}(\textbf{V}) = 1$, the optimal reflection matrix $\mathbf{\Theta}^{*}$ can be derived by calculating the eigenvalue and eigenvector of $\textbf{V}$.
When $\text{rank}(\textbf{V}) \ne 1$, the GR method is adopted, and the eigenvalue decomposition of $\textbf{V}$ is defined as
\begin{eqnarray}
\label{eigenvalue}
\mathbf{V} = \mathbf{U} \mathbf{\Sigma} \mathbf{U}^{H},
\end{eqnarray}
where $\mathbf{U} = [e_1,e_2, \ldots, e_{M+1}]$ is a unitary matrix of eigenvectors, and $\mathbf{\Sigma} = \text{diag} \{ \varpi_1,\varpi_2, \ldots, \varpi_{M+1} \}$ is a diagonal matrix of eigenvalues.

\begin{spacing}{1.3}
Then, we generate two independent normally distributed random vectors $\mathbf{x} \in \mathbb{R}^{(M+1) \times 1}$ and $\mathbf{y} \in \mathbb{R}^{(M+1) \times 1}$ with zero mean and covariance matrix $\frac{1}{2} \mathbf{I}_{M+1}$.
Let $N$ denote the maximum generation of candidate random vectors, and the Gaussian random vector in the $n$-th generation is given by
\begin{eqnarray}
\label{random_vector}
\mathbf{r}_{n} = \mathbf{x} + \mathbf{y} \sqrt{-1}, \ n =1,2, \ldots, N.
\end{eqnarray}

Based on the generated Gaussian random vector $\mathbf{r}_{n} \in \mathcal{CN} (\mathbf{0}, \mathbf{I}_{M+1}) $ in the complex plane, we can obtain a suboptimal solution to (\ref{problem_feasibility_check_theta_abcd}), denoting as 
\begin{eqnarray}
\label{suboptimal}
\boldsymbol{\bar{\nu}}_{n} = \mathbf{U} \mathbf{\Sigma}^{1/2} \mathbf{r}_{n}, \ n =1,2, \ldots, N.
\end{eqnarray}

Next, the candidate reflection matrix can be expressed as
\begin{eqnarray}
\label{refelction}
\mathbf{\Theta}_{n} = \text{diag} \left\lbrace e^{j \arg( \frac{ \boldsymbol{\bar{\nu}}_{n}[m] }{ \boldsymbol{\bar{\nu}}_{n}[M+1] } ) } \ | \ \forall m \in \mathcal{M} \right\rbrace,
\end{eqnarray}
where $\boldsymbol{\bar{\nu}}_{n}[m]$ denotes the $m$-th elements of $\boldsymbol{\bar{\nu}}_{n}$.
With the obtained candidate set of reflection matrix $\{ \mathbf{\Theta}_{n} | n =1,2, \ldots, N \}$, we can find the optimal one that maximizes the combined channel gains of all users, i.e.,
\begin{eqnarray}
\label{argmax}
n^* = \arg \max_{n} \sum \nolimits_{i} \sum \nolimits_{j} \sum \nolimits_{k} |h_{ijk} + \mathbf{g}_{ik}^{H} \mathbf{\Theta}_{n} \mathbf{f}_{jk}|^2.  
\end{eqnarray}
\end{spacing}


%
Based on the searched reflection matrix $\mathbf{\Theta}^{*}$, if the combined channel gains experienced by any two users $(i, \tilde{i})$ associated with BS $j$ on subchannel $k$ can be arranged as $H_{ijk} \le H_{\tilde{i}jk}$, then the decoding order is given by $\pi_{jk}(i) \le \pi_{jk}(\tilde{i})$.
According to the above discussions, the co-design method for reflection matrix and decoding order can be summarized in \textbf{Algorithm \ref{codesign_of_reflection_decoding}}.

\begin{algorithm}
	\caption{Co-design of Reflection Matrix and Decoding Order}
	\label{codesign_of_reflection_decoding}
	\begin{algorithmic}[1]
		\renewcommand{\algorithmicrequire}{\textbf{Initialize}}
		\renewcommand{\algorithmicensure}{\textbf{Output}}
		\STATE \textbf{Initialize} the maximum generation of candidate random vector as $N_{4}$, solve the relaxed SDP problem of (\ref{problem_feasibility_check_theta_abcd}) and obtain an optimal solution $\mathbf{V}$.
		\IF{$\text{rank}(\mathbf{V}) = 1$}
		\STATE With obtained $\mathbf{V}$, calculate its eigenvalue $\varpi$ and eigenvector $\mathbf{u}$ according to $\mathbf{V} \mathbf{u} = \varpi \mathbf{u} $;
		\STATE Update $\mathbf{\Theta}^{*} := \text{diag} \{ \sqrt{\varpi} \mathbf{u} \} $;
		\ELSE
		\STATE Obtain the eigenvalue decomposition using (\ref{eigenvalue});
		\FOR{$n_{4}=1,2, \ldots, N_{4}$}
		\STATE Generate a Gaussian random vector $\mathbf{r}_{n_{4}}$ using (\ref{random_vector});
		\STATE Obtain a candidate solution $\mathbf{\Theta}_{n_{4}}$ using (\ref{suboptimal}) and (\ref{refelction});
		\ENDFOR
		\STATE Find the optimal $\mathbf{\Theta}^{*} := \mathbf{\Theta}_{n_{4}^*}$ according to (\ref{argmax});
		\ENDIF
		\STATE With the optimal $\mathbf{\Theta}^{*}$, calculate all combined channel gains $\{ |h_{ijk} + \mathbf{g}_{ik}^{H} \mathbf{\Theta}^{*} \mathbf{f}_{jk}|^2 \ | \ \forall j,k \}$ and rank them in ascending order for each BS $j$ on subchannel $k$;
		\STATE \textbf{Output} the optimal reflection matrix $\mathbf{\Theta}^{*}$ and decoding order $\pi_{jk}^{*}, \ \forall j,k$.
	\end{algorithmic}
\end{algorithm}

\subsection{Convergence and Complexity Analysis} \label{section_3_analysis}
\subsubsection{Convergence}
{In \textbf{Algorithm \ref{algorithm_1}}, we denote $\mathbf{p}^{(n_{2})}$ and $\boldsymbol{\gamma}^{(n_{2})}$ as the solution of problem (\ref{problem_power_lambda}) obtained at the $n_{2}$-th iteration, where the utility value is given by $U^{(n_{2})} = U \left( \boldsymbol{\gamma}^{(n_{2})} \right) $.
Then, the coefficient sets $\boldsymbol{\lambda}^{(n_{2}+1)}$ and $\boldsymbol{\bar{\lambda}}^{(n_{2}+1)}$ can be updated by (\ref{CUB_update_lambda_1}) and (\ref{CUB_update_lambda_2});
Note that the utility value in \textbf{Algorithm \ref{algorithm_1}} only depends on $\boldsymbol{\gamma}$, such that
\begin{equation}
\label{equation_30}
	U \left( \boldsymbol{\gamma}^{(n_{2})} \right) = U \left( \mathbf{p}^{(n_{2})}, \boldsymbol{\gamma}^{(n_{2})} \right) = U \left( \mathbf{p}^{(n_{2})}, \boldsymbol{\gamma}^{(n_{2})}, \boldsymbol{\lambda}^{(n_{2}+1)}, \boldsymbol{\bar{\lambda}}^{(n_{2}+1)} \right).
\end{equation}

By substituting $\boldsymbol{\lambda}^{(n_{2}+1)}$ and $\boldsymbol{\bar{\lambda}}^{(n_{2}+1)}$ into the problem (\ref{problem_power_lambda}), we can obtain $\mathbf{p}^{(n_{2}+1)}$ and $\boldsymbol{\gamma}^{(n_{2}+1)}$ by solving the resulting problem once again, and thus we have
\begin{equation} \label{equation_31}
U \left( \mathbf{p}^{(n_{2})}, \boldsymbol{\gamma}^{(n_{2})}, \boldsymbol{\lambda}^{(n_{2}+1)}, \boldsymbol{\bar{\lambda}}^{(n_{2}+1)} \right)
\le
U \left( \mathbf{p}^{(n_{2}+1)}, \boldsymbol{\gamma}^{(n_{2}+1)}, \boldsymbol{\lambda}^{(n_{2}+1)}, \boldsymbol{\bar{\lambda}}^{(n_{2}+1)} \right).
\end{equation}

Similar to (\ref{equation_30}), the utility value only depends on $\boldsymbol{\gamma}$.
Thus, the following equation is satisfied:}
\begin{equation} \label{equation_32}
	\begin{aligned}
	U \left( \boldsymbol{\gamma}^{(n_{2}+1)} \right)
	& = U \left( \mathbf{p}^{(n_{2}+1)}, \boldsymbol{\gamma}^{(n_{2}+1)} \right)
	= U \left( \mathbf{p}^{(n_{2}+1)}, \boldsymbol{\gamma}^{(n_{2}+1)}, \boldsymbol{\lambda}^{(n_{2}+2)}, \boldsymbol{\bar{\lambda}}^{(n_{2}+2)} \right) \\
	& = U \left( \mathbf{p}^{(n_{2}+1)}, \boldsymbol{\gamma}^{(n_{2}+1)}, \boldsymbol{\lambda}^{(n_{2}+1)}, \boldsymbol{\bar{\lambda}}^{(n_{2}+1)} \right)
	\end{aligned}
\end{equation}

Therefore, combining $(\ref{equation_30})$, $(\ref{equation_31})$ and $(\ref{equation_32})$, it can be observed that the utility value of problem (\ref{problem_power_lambda}) is non-decreasing over iterations, which can be expressed as
\begin{equation}
U^{(n_{2})} = U \left( \boldsymbol{\gamma}^{(n_{2})} \right)
\le
U \left( \boldsymbol{\gamma}^{(n_{2}+1)} \right) = U^{(n_{2}+1)}.
\end{equation}

Finally, due to the fact that the system bandwidth and available transmission power are limited in practice, the utility value (i.e., achievable sum rate) has an upper bound.
Hence, \textbf{Algorithm \ref{algorithm_1}} is guaranteed to converge as long as the value of $N_{2}$ is set large enough.
The convergence proofs of \textbf{Algorithm \ref{algorithm_feasibility_searching}} is omitted here for brevity, due to their similar derivations.

\subsubsection{Complexity}
{When the convex problems are solved by CVX, the interior point method is considered, unless otherwise stated. 
In \textbf{Algorithm \ref{algorithm_1}}, the dimension of variables to be solved is $2IJK$. Thus, the complexity of \textbf{Algorithm \ref{algorithm_1}} can be expressed as $\mathcal{O} \left( N_2 (2IJK)^3 \right)$, where $N_2$ is the maximal iteration number for finding the converged power allocation strategy.
Similarly, the complexity of \textbf{Algorithm \ref{algorithm_feasibility_searching}} is bounded by $\mathcal{O} \left( N_3 (2IJK)^3 \right)$.
In \textbf{Algorithm \ref{codesign_of_reflection_decoding}}, the complexity for solving the relaxed SDP problem of (\ref{problem_feasibility_check_theta_abcd}) is $\mathcal{O} \left( (M+4IJK)^6 \right)$.
Meanwhile, define $N_4$ as the maximal number of the generated Gaussian random vectors, and denote $T_{GR}$ as the complexity of performing one Gaussian random.
Thereby, the complexity of \textbf{Algorithm \ref{codesign_of_reflection_decoding}} can be expressed as $\mathcal{O} \left( (M+4IJK)^6 + N_4 T_{GR} \right)$ in the worst case.}

\section{Matching Theory for User Association \\ and Subchannel Assignment}
\label{matching}
In this section, we focus on the user association and subchannel assignment problem in (\ref{max_sum_rate}) with fixed power allocation and reflection matrix, which can be expressed as
\begin{subequations}
	\label{3D_matching}
	\begin{eqnarray}
	\label{matching_problem}
	&\max \limits_{\boldsymbol{\alpha},\boldsymbol{\beta}} & \sum_{i=1}^{I} \sum_{j=1}^{J} \sum_{k=1}^{K} R_{ijk}, \\
	&{\rm s.t.}& {\rm (7f)-(7j)},
	\end{eqnarray}
\end{subequations}
where $\boldsymbol{\alpha}=\{\alpha_{ij} | \forall i,j\}$ denotes the user association profile and $\boldsymbol{\beta} = \{ \beta_{jk} | \forall j,k\}$ represents the subchannel assignment profile.
It can be observed that (\ref{3D_matching}) is a 3D matching problem involving three finite and disjoint sets (i.e., user set $\mathcal{I}$, BS set $\mathcal{J}$, and subchannel set $\mathcal{K}$), which is proved to be NP-hard for obtaining the optimal solution.
In order to address this challenging issue, we decompose the 3D matching problem (\ref{3D_matching}) into two 2D matching problems, namely, the user association problem and subchannel assignment problem.
The former problem is to cluster all users into multiple disjoint user groups, and the latter problem is to assign all subchannels into multiple subchannel sets.
More expectantly, in the user association problem, the users in each group form a cell served by one BS through the NOMA transmission.
Thus, it is a many-to-one matching problem.
In the subchannel assignment problem, one subchannel can be reused by multiple BSs and multi-subchannel can be assigned to one BS, which is a many-to-many matching problem.

\subsection{Matching Problem Formulation}
Before solving the aforementioned two matching problems, we give the following remarks and definitions for ease of exposition.

\begin{remark}
	\emph{The above mentioned 2D matching problems is a many-to-many (one) matching problem with peer effects.}
\end{remark}

\begin{IEEEproof}
	On the one hand, in the user association problem, owing to the feature of multiplexing power domain NOMA, the achievable data rate of any user $i$ associated with BS $j$ over all subchannels is related to other paired users sharing the same subchannel.
	As a result, each BS should take into account the internal relationship of the associated users when it selects a certain user to match with.
	It is the intra-cell interference that makes the problem of user association a many-to-one matching problem with peer effects.
	On the other hand, in the subchannel assignment problem, owing to the reuse of subchannels among different cells, the sum rate of each BS $j$ over subchannel $k$ is affected by other BS assigned with the same subchannel.
	Thus, the preference of each BS not only depends on the subchannel it matches with, but also depends on other BSs that match with the same subchannel.
	Therefore, the individual BS preference depends on other peers, and it is the inter-cell interference that makes the problem of subchannel assignment a many-to-many matching problem with peer effects, which completes the proof.
\end{IEEEproof}

\begin{definition} [2D Matching]
	A matching $\mu$ is a function from the set $\mathcal{E} \bigcup \mathcal{W}$ to the set of all subsets of $\mathcal{E} \bigcup \mathcal{W}$ such that
	1) $\mu(e) \subseteq \mathcal{W}$ and $|\mu(e)|=\ell_w $, $\forall e \in \mathcal{E}$;
	2) $\mu(w) \subseteq \mathcal{E}$ and $|\mu(w)|=\ell_e $, $\forall w \in \mathcal{W}$;
	3) $\mu(e) \subseteq \mathcal{W}$ if and only if $\mu(w) \subseteq \mathcal{E}$;
	4) $e \in \mu(w)$ \ if and only if $w \in \mu(e)$;
	where $\mathcal{E} = \{e_1,\ldots,e_n\}$ and $\mathcal{W} = \{w_1,\ldots,w_u\}$ are two finite and disjoint player sets, $\ell_w$ and $\ell_e$ are two positive integers.
\end{definition}

Note that the above condition 1) implies that each player $e \in \mathcal{E}$ can be matched with $\ell_w$ players in $\mathcal{W}$.
Similarly, condition 2) means that each player $w \in \mathcal{W}$ can be matched with $\ell_e$ players in $\mathcal{E}$.
Condition 3) indicates that the mapping of player $e \in \mathcal{E}$ is the subset of $\mathcal{W}$, and vice versa.
Condition 4) represents that if player $e \in \mathcal{E}$ matched with $w \in \mathcal{W}$, then player $w \in \mathcal{W}$ is also matched with $e \in \mathcal{E}$.
It is worth noting that when $\ell_w \ge 2$ and $\ell_e \ge 2$, then one can obtain the definition of many-to-many matching.
When $\ell_w \ge 2$ and $\ell_e = 1$, it becomes a many-to-one matching.

\begin{remark}
	\emph{The formulated many-to-many (one) matching problem is lack of the property of substitutability.}
\end{remark}

\begin{IEEEproof}
	Given player set $\mathcal{E}$ and $\mathcal{W}$, each player $e \in \mathcal{E}$ can determine which subset of $\mathcal{W}$ it is most likely to match with.
	This is called the choice set of $e$ in $\mathcal{W}$, denoted by $C_e(\mathcal{W}) = \mathcal{W}'$.
	That is, the player $e$ prefers $\mathcal{W}'$ to any subset of $\mathcal{W}$, which can be expressed as
	\begin{equation}
	\forall \ \mathcal{W}'' \subset \mathcal{W}, \ \mathcal{W}'' \ne \mathcal{W}' \Rightarrow \mathcal{W}' \succ_{e} \mathcal{W}''.
	\end{equation}
	
	For any set $\mathcal{W}$ that contains $w$ and $w'$, the preference of $e$ over sets of $\mathcal{W}$ has the property of substitutability if and only if $w \in C_e(\mathcal{W})$ and $w \in C_e(\mathcal{W} \backslash \{w'\})$.
	It means that when a player $e \in \mathcal{E}$ has the property of substitutability, it regards the players in the choice set $C_e(\mathcal{W})$ as alternatives rather than complements, even if a player $w' \in \mathcal{W}$ in the choice set rejects it, its selection of other players in the choice set will not be affected.
	Nevertheless, on the one hand, due to the intra-cell interference from the user pairing in the user association problem, the achievable data rate of user $i$ associated with BS $j$ may change after its paired user $i'$ is unmatched with BS $j$.
	Thus, user $i$ may not be in the preferred set of BS $j$ any more, which implies that the formulated many-to-one user association problem does not have the property of substitutability.
	On the other hand, due to the inter-cell interference among BSs assigned with the same subchannels, the achievable rate of subchannel $k$ with BS $j$ may change after $j'$ is unmatched with $k$.
	Hence, BS $j$ may not be in the preferred set of subchannel $k$ any more, which indicates that the formulated many-to-many subchannel assignment problem does not have the property of substitutability as well, which completes the proof.
\end{IEEEproof}

During the matching process, each player $e \in \mathcal{E}$ has a transitive and strict preference list w.r.t. its interests over the set of $\mathcal{W}$, and vice versa.
We use $w_1 \succ_{e} w_2$ to denote that player $e$ strictly prefers $w_1$ to $w_2$.
If $w_2 \succ_{e} w_3$ is satisfied at the same time, then we have $w_1 \succ_{e} w_3$.
Due to the existence of peer effects and non-substitutability in the formulated many-to-many (one) matching problem, the preference lists of players vary continuously over the matching process, which makes the matching mechanisms complex to design.
Given a matching function $\mu$, and assume that $\mu(e)=w$ and $\mu(e')=w'$.
Then, in order to handle the peer effects and ensure exchange stability, we define the swap matching as
\begin{equation}
\label{swap_matching}
\mu_{e}^{e'} = \left\lbrace \mu \backslash \{ (e,w),(e',w')\} \bigcup \{(e',w),(e,w')\} \right\rbrace,
\end{equation}
where players $e$ and $e'$ exchange their matched players $w$ and $w'$ while keeping all other matching states the same.
Based on the swap operation in (\ref{swap_matching}), we define the concept of swap-blocking pair as follows.

\begin{definition}[Swap-Blocking Pair]
	\label{swap_operation}
	A pair of players $(e,e')$ is called a swap-blocking pair in $\mu$ if and only if 
	1) $\forall q \in \{e,e',w,w'\}$, $U_q(\mu_{e}^{e'}) \ge U_q(\mu)$;
	2) $\exists q \in \{e,e',w,w'\}$, such that $U_q(\mu_{e}^{e'}) > U_q(\mu)$;
	where $U_q(\mu)$ denotes the utility of player $q$ under matching $\mu$.
\end{definition}

The aforementioned condition 1) shows that the utilities of all involved players should not be decreased after the swap operation.
Condition 2) implies that at least one of the involved payer's utilities is increased after the swap operation.
What is worth mentioning is that the matching $\mu$ is two-sided exchange-stable if and only if there dose not exist a swap-blocking pair. 
Otherwise, the swap matching $\mu_{e}^{e'}$ in a swap-blocking pair would be approved, and the achievable utilities of the involved players will not decrease and at least one player's utility will increase after the swap operation.


\subsection{Many-to-One Matching for User Association}
\label{5-A}
In the many-to-one matching problem of user association, we define the preference of each user $i$ associated with BS $j$ as
\begin{equation}
U_{ij} = \sum \limits_{k \in \mathcal{K}} \frac{W}{K} \log_{2} \left( 1+\gamma_{ijk} \right).
\end{equation}

If user $i$ can achieve a higher data rate when being associated with BS $j$ compared to be that of being associated with BS $j'$, i.e., user $i$ prefers to be associated with the BS $j$ in matching $\mu$ rather than the BS $j'$ in matching $\mu'$, then we have 
\begin{equation}
\label{user_preference}
(j,\mu) \succ_{i} (j',\mu') \ \Leftrightarrow \ U_{ij}(\mu) > U_{ij'}(\mu').
\end{equation}

Similarly, the preference of each BS $j$ associated with a set of users $\mu(j)$ can be given by
\begin{equation}
U_{j} = \sum \limits_{i \in \mu(j)} \sum \limits_{k \in \mathcal{K}} \frac{W}{K} \log_{2} \left( 1+\gamma_{ijk} \right).
\end{equation}

For any two subsets of users $\mathcal{I}_1=\mu(j)$ and $\mathcal{I}_2=\mu'(j)$ while $\mathcal{I}_1 \ne \mathcal{I}_2$, if BS $j$ obtain get a higher data rate when being associated with $\mathcal{I}_1$ than that of being associated to $\mathcal{I}_2$, i.e., BS $j$ prefers the user subset $\mathcal{I}_1$ in matching $\mu$ to the user subset $\mathcal{I}_2$ in matching $\mu'$, then we have
\begin{equation}
\label{BS_preference}
(\mathcal{I}_1,\mu) \succ_{j} (\mathcal{I}_2,\mu') \ \Leftrightarrow \ U_{j}(\mu) > U_{j}(\mu').
\end{equation}

According to (\ref{user_preference}) and (\ref{BS_preference}), the preference lists of all users and BSs are constructed.
Subsequently, each user proposes to the most preferred BS that has never rejected them before.
Then, each BS accepts the most preferred users and rejects the others.
Finally, the initial matching state between users and BSs is obtained when there is no unmatched user.
After that, each user tries to search for another user to form a swap-blocking pair and swaps their matching states based on (\ref{swap_matching}), which terminates when no swap-blocking pair exists.
In summary, the many-to-one matching for user association is described in \textbf{Algorithm \ref{algorithm_4}}.

\begin{algorithm}[h]
	\caption{Many-to-One Matching for User Association}
	\label{algorithm_4}
	\begin{algorithmic}[1]
		\renewcommand{\algorithmicrequire}{\textbf{Initialize}}
		\renewcommand{\algorithmicensure}{\textbf{Output}}
		\STATE \textbf{Initialize} the User-BS matching state as $\Phi_{1}$.
		\REPEAT
		\STATE For every user $i \in \Phi_{1}$, it searches for another user $i' \in \Phi_{1} \backslash \Phi_{1}(\mu(i))$ to check whether $(i,i')$ is a swap-blocking pair;
		\IF {$(i,i')$ is a swap-blocking pair}
		\STATE Update $\mu := \mu_{i}^{i'}$;
		\ELSE
		\STATE Keep the current matching state;
		\ENDIF
		\UNTIL No swap-blocking pair can be constructed.
		\STATE \textbf{Output} the stable User-BS matching $\mu^{*}$ and its corresponding utility $U_1=U(\mu^{*})$.
	\end{algorithmic}
\end{algorithm}

\subsection{Many-to-Many Matching for Subchannel Assignment}
Analogously, in the many-to-many matching problem of subchannel assignment, the preference of each BS $j$ assigned with subchannel $k$ is defined as
\begin{equation}
U_{jk} = \sum \limits_{i \in \mathcal{I}} \frac{W}{K} \log_{2} \left( 1+\gamma_{ijk} \right).
\end{equation}

If BS $j$ can achieve a higher data rate when being assigned with subchannel $k$ compared to that of being assigned with subchannel $k'$, i.e., BS $j$ prefers to the subchannel $k$ in matching $\mu$ rather than the subchannel $k'$ in matching $\mu'$, then we have 
\begin{equation}
\label{user_BS_preference}
(k,\mu) \succ_{j} (k',\mu') \ \Leftrightarrow \ U_{jk}(\mu) > U_{jk'}(\mu').
\end{equation}

Similarly, the preference of each subchannel $k$ on a set of BSs $\mu(k)$ can be given by
\begin{equation}
U_{k} = \sum \limits_{i \in \mathcal{I}} \sum \limits_{j \in \mu(k)} \frac{W}{K} \log_{2} \left( 1+\gamma_{ijk} \right).
\end{equation}

For any two subsets of BSs $\mathcal{J}_1=\mu(k)$ and $\mathcal{J}_2=\mu'(k)$ while $\mathcal{J}_1 \ne \mathcal{J}_2$, if subchannel $k$ can get a higher data rate when being assigned to $\mathcal{J}_1$ than $\mathcal{J}_2$, i.e., subchannel $k$ prefers to the BS subset $\mathcal{J}_1$ in matching $\mu$ rather than the BS subset $\mathcal{J}_2$ in matching $\mu'$, then we have
\begin{equation}
\label{subchannel_preference}
(\mathcal{J}_1,\mu) \succ_{k} (\mathcal{J}_2,\mu') \ \Leftrightarrow \ U_{k}(\mu) > U_{k}(\mu').
\end{equation}

First, the preference lists of all (User-BS) units and subchannels are established according to (\ref{user_BS_preference}) and (\ref{subchannel_preference}).
Then, an initial matching state can be generated by adopting the aforementioned method in Section \ref{5-A}.
Finally, the search process is executed based on (\ref{swap_matching}), which terminates until there exists no swap-blocking pair.
The many-to-many matching for subchannel assignment is described in \textbf{Algorithm \ref{algorithm_5}}.

\begin{algorithm}[h]
	\caption{Many-to-Many Matching for Subchannel Assignment}
	\label{algorithm_5}
	\begin{algorithmic}[1]
		\renewcommand{\algorithmicrequire}{\textbf{Initialize}}
		\renewcommand{\algorithmicensure}{\textbf{Output}}
		\STATE \textbf{Initialize} the (User,BS)-Subchannel matching state as $\Phi_{2}$. 
		\REPEAT
		\STATE For every (User,BS) $j \in \Phi_{2}$, it searches for another (User,BS) $j' \in \Phi_{2} \backslash \Phi_{2}(\mu(j))$, and let $\mathcal{U}=\{U_1\}$;
		\STATE For a given $j$, calculate the candidate $U_{j}^{j'}$ for the swapping pair $(j,j')$;
		\IF {$(j,j')$ is a swap-blocking pair}
		\STATE Update $\mathcal{U} := \mathcal{U} \cup \{U_{j}^{j'}\}$;
		\ENDIF
		\STATE Find $j^{'*} = \arg \max_{j'} \mathcal{U}$;
		\STATE Update $\bar{\mu} := \bar{\mu}_{j}^{j^{'*}}$, and set $U_2 = U_{j}^{j^{'*}}$;
		\UNTIL No swap-blocking pair can be constructed.
		\STATE \textbf{Output} the stable (User,BS)-Subchannel matching $\bar{\mu}^{*}$.
	\end{algorithmic}
\end{algorithm}

\subsection{Property Analysis} \label{section_4_analysis}
The properties in terms of stability, convergence, complexity and optimality of the proposed matching-based algorithms are analyzed in the following propositions.
\begin{proposition}[Stability]
	\emph{The final matching $\mu^{*}$ and $\bar{\mu}^{*}$ derived from \textbf{Algorithm \ref{algorithm_4}} and \textbf{Algorithm \ref{algorithm_5}} are both two-sided exchange-stable matching.}
\end{proposition}

\begin{IEEEproof}
	This proposition can be proved by contradiction.
	Assume that there exists a blocking pair ($i$,$i'$) in the final matching $\mu^{*}$ satisfying that $\forall q \in \{ i,i',\mu(i), \mu(i') \}$, $U_q \left( (\mu^{*})_i^{i'} \right) \ge U_q \left( \mu^{*} \right)$ and $\exists q \in \{ i,i',\mu(i), \mu(i') \}$ such that $U_q \left( (\mu^{*})_i^{i'} \right) > U_q \left( \mu^{*} \right)$.
	According to step 2 to step 9 in \textbf{Algorithm \ref{algorithm_4}}, the swap operation continues until there exists no swap-blocking pair.
	That is to say, $\mu^{*}$ is not the final matching, which contradicts our initial assumption and the proposition is proved.
	As a result, it can be concluded that the proposed algorithm reaches a two-sided exchange stability in the end.
	The proof for $\bar{\mu}^{*}$ in \textbf{Algorithm \ref{algorithm_5}} can be derived similarly, which is omitted here for brevity.
\end{IEEEproof}

\begin{proposition}[Convergence]
	\emph{Both \textbf{Algorithm \ref{algorithm_4}} and \textbf{Algorithm \ref{algorithm_5}} converge to a two-sided exchange-stable matching within a limited number of iterations.}
\end{proposition}

\begin{IEEEproof}
	Given a matching function $\mu$ for the user association problem, suppose that $\mu(i)=j$, $\mu(i')=j'$, while $(i,i')$ is a swap-blocking pair.
	According to \textbf{Definition \ref{swap_operation}}, at least one of the utilities of BS $j$ and $j'$ increases after the swap operation.
	Thus, there are three cases:
	\rmnum{1}) $U_{j}(\mu_i^{i'}) > U_j(\mu)$ and $U_{j'}(\mu_i^{i'}) > U_{j'}(\mu)$;
	\rmnum{2}) $U_{j}(\mu_i^{i'}) = U_j(\mu)$ and $U_{j'}(\mu_i^{i'}) > U_{j'}(\mu)$;
	\rmnum{3}) $U_{j}(\mu_i^{i'}) > U_j(\mu)$ and $U_{j'}(\mu_i^{i'}) = U_{j'}(\mu)$.
	It can be observed that the utilities of the involved BSs are non-decreasing, and the achievable sum rate of each BS has an upper bound due to the limited system bandwidth and transmission power constraint in practice.
	Therefore, the number of iterations of \textbf{Algorithm \ref{algorithm_4}} is limited, and it converges to a two-sided exchange-stable matching when there exists no swap-blocking pair that can further improve any player's utility.
	The convergence proof for \textbf{Algorithm \ref{algorithm_5}} can be derived similarly, which is omitted here for brevity.
\end{IEEEproof}

\begin{proposition}[Complexity]
	\emph{The computational complexity of \textbf{Algorithm \ref{algorithm_4}} and \textbf{Algorithm \ref{algorithm_5}} is upper bounded by $\mathcal{O}(IJ^2 + A_{\text{max}}IJN_{it})$ and $\mathcal{O}\left( J^2 K^2(\bar{N}_{it}+1)\right)$, respectively.}
\end{proposition}

\begin{IEEEproof}
	The complexity of the proposed matching-based algorithms depends on the initialization and swap process.
	In \textbf{Algorithm \ref{algorithm_4}}, the initialization process requires each user to propose to one BS and each BS can accept or reject the proposal based on its preference.
	The complexity of constructing the initial User-BS matching state is $\mathcal{O}(IJ^2)$ in the worst case.
	For the swap process in \textbf{Algorithm \ref{algorithm_4}}, there are no more than $A_{\text{max}}$ users in each cell can perform the swap operation with other $(J-1)$ unassociated BSs, and thus the maximum swap operation number for each user is $A_{\text{max}}(J-1)$.
	Let $N_{it}$ denote the number of total iteration when there is no swap-blocking pair.
	Thus, the complexity of swap operation is $\mathcal{O}(A_{\text{max}}IJN_{it})$.
	Overall, the complexity of \textbf{Algorithm \ref{algorithm_4}} can be calculated as $\mathcal{O}(IJ^2 + A_{\text{max}}IJN_{it})$.
	
	In \textbf{Algorithm \ref{algorithm_5}}, each (User-BS) unit can propose to multiple subchannels and each subchannel decides to accept or reject the proposal based on its preference.
	The complexity of the initialization process in \textbf{Algorithm \ref{algorithm_5}} is $\mathcal{O}(J^2 K^2)$ in the worst case.
	For the swap process in \textbf{Algorithm \ref{algorithm_5}}, each (User-BS) unit can perform the swap operation with other $(J-1)$ units for a given subchannel, each subchannel can perform the swap operation with other $(K-1)$ subchannels for a given (User-BS) unit.
	Let $\bar{N}_{it}$ denote the number of total iteration when there is no swap-blocking pair.
	Thus, the complexity of swap operation is $\mathcal{O}\left( J(J-1)K(K-1)\bar{N}_{it}\right) $.
	Overall, the complexity of \textbf{Algorithm \ref{algorithm_5}} can be calculated as $\mathcal{O}\left( J^2 K^2(\bar{N}_{it}+1)\right) $, which completes the proof.
\end{IEEEproof}

\begin{proposition}[Optimality]
	\emph{All local optimal utilities of \textbf{Algorithm \ref{algorithm_4}} and \textbf{Algorithm \ref{algorithm_5}} correspond to a two-sided exchange stable matching, but not vice versa.}
\end{proposition}

\begin{IEEEproof}
	This proposition can be proved by contradiction.
	Suppose that the converged utility $U_1=U(\mu)$ of \textbf{Algorithm \ref{algorithm_4}} is a local optimal value.
	If $\mu$ is not a two-sided exchange stable matching, it means that we can find a swap-blocking pair to further improve the utilities of users and/or BSs, which contradicts our initial assumption that the utility $U_1=U(\mu)$ is a local optimum.
	Therefore, it can be concluded that $\mu$ is a two-sided exchange stable matching.
	However, not all two-sided exchange stable matchings $\mu$ correspond to a local optimum of utility.
	This can be explained by the following example:
	given a stable matching $\mu$, and assume that $j=\mu(i)$, $j'=\mu(i')$.
	It can be observed that $(i,i')$ is not a swap-blocking pair when $\mu$ is a stable matching. Thus, BS $j$ will not approve a swap operation with BS $j'$, due to the fact that none of BSs' utilities is improved after the swap operation.
	But, user $i$ and $i'$ will reap a lot of benefits if this swap operation is accepted, which may further improve the utility of BSs.
	The proof for \textbf{Algorithm \ref{algorithm_5}} can be derived similarly, which is omitted here for brevity.
\end{IEEEproof}

\section{Numerical Results}
\label{simulation}

\label{settings}
We consider that there are $6$ users, $3$ BSs and $3$ subchannels in the IRS-aided NOMA network.
Specifically, in the 3D Cartesian coordinates, the location of user $i$ is denoted by $(x_i,y_i,z_i) = (50i,30,0)$, $i=1,2,\ldots, 6$,
the location of BS $j$ is denoted by $(x_j,y_j,z_j) = (100j,0,20)$, $j=1,2,3$,
and the location of the IRS is denoted by $(x_{\text{IRS}},y_{\text{IRS}},z_{\text{IRS}}) =(200,50,20)$.
We assume that the path loss model is given by $L(d) = \varsigma_0 (d)^{-a} $, where $\varsigma_0 = -30$ dB is the path loss at the reference distance of 1 meter, $d$ denotes the link distance, and $a$ is the path loss exponent.
{Specifically, the path loss exponent of the BS-user, IRS-user and BS-IRS links are set as 3.2, 2.6, and 2.2, respectively \cite{Hou2020Reconfigurable, Zheng2020IRS}.}
The small-scale fading model is given by ${F} = \sqrt{\frac{\kappa}{1+\kappa}} {F}^{\text{LoS}} + \sqrt{\frac{1}{1+\kappa}} {F}^{\text{NLoS}}$, where $\kappa = 2$ is the Rician factor for the BS-IRS link, ${F}^{\text{LoS}}$ denotes the deterministic line-of-sight (LoS) channel component with $|{F}^{\text{LoS}}| = 1$ and ${F}^{\text{NLoS}}$ is random non-line-of-sight (NLoS) channel component that follows the Rayleigh distribution with parameter $\iota=1$.
In particular, the small-scale fading ${F}$ is simplified to Rayleigh fading when $\kappa=0$, which is applicable for the BS-user and IRS-user links.
Then, the channel gain equals to the small-scale fading multiplied by the square root of the path loss.
Moreover, the number of reflecting elements is set as $M=100$, the system bandwidth is assumed to be $W=3$ MHz.
The noise power is $\sigma^2 = -80$ dBm, and the minimum rate requirement of each user is assumed to be $R_{\text{min}}=500$ Kbps.
The maximum transmission power of each BS is set as $P_{\text{max}}=23$ dBm, unless otherwise stated.

In order to validate the effectiveness of our proposed algorithms for the IRS-aided multi-cell NOMA network with multiple subchannels, the following three schemes are considered as benchmarks:
1) {OMA without IRS}: 
Frequency reuse and time division multiple access (TDMA) are considered in a multi-cell OMA network, where an BS communicates with at most one user in each time slot;
2) {OMA with IRS}: Compared to scheme 1, the only difference in scheme 2 is that there is one passive IRS with finite reflecting elements whose reflection matrix can be adjusted to intelligently reconfigure the wireless communication environment;
3) {NOMA without IRS}: All frequency can be reused by adjacent cells, and the SIC approach is applied in each cell to decode the intended signal of each user.
Furthermore, we simulate 2000 trials, and all results are averaged over independent channel realization.

\begin{figure*}[t]
	\begin{minipage}[t]{0.3 \textwidth}
		\centering
		\includegraphics[width=2.3 in]{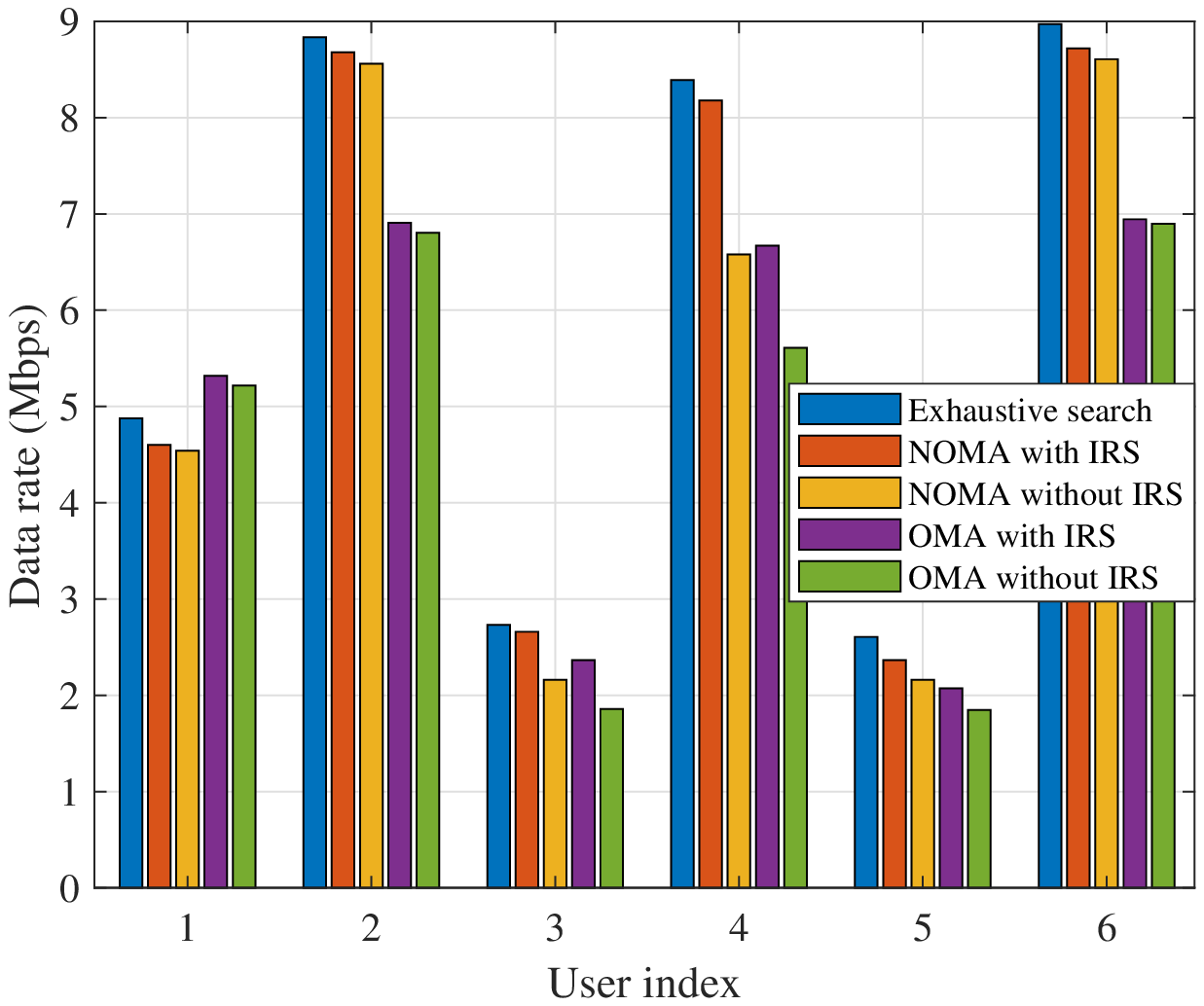}
		\caption{Individual data rate.}
		\label{ES_individual_data_rate}
	\end{minipage}
	\hspace{2 mm}
	\begin{minipage}[t]{0.3 \textwidth}
		\centering
		\includegraphics[width=2.3 in]{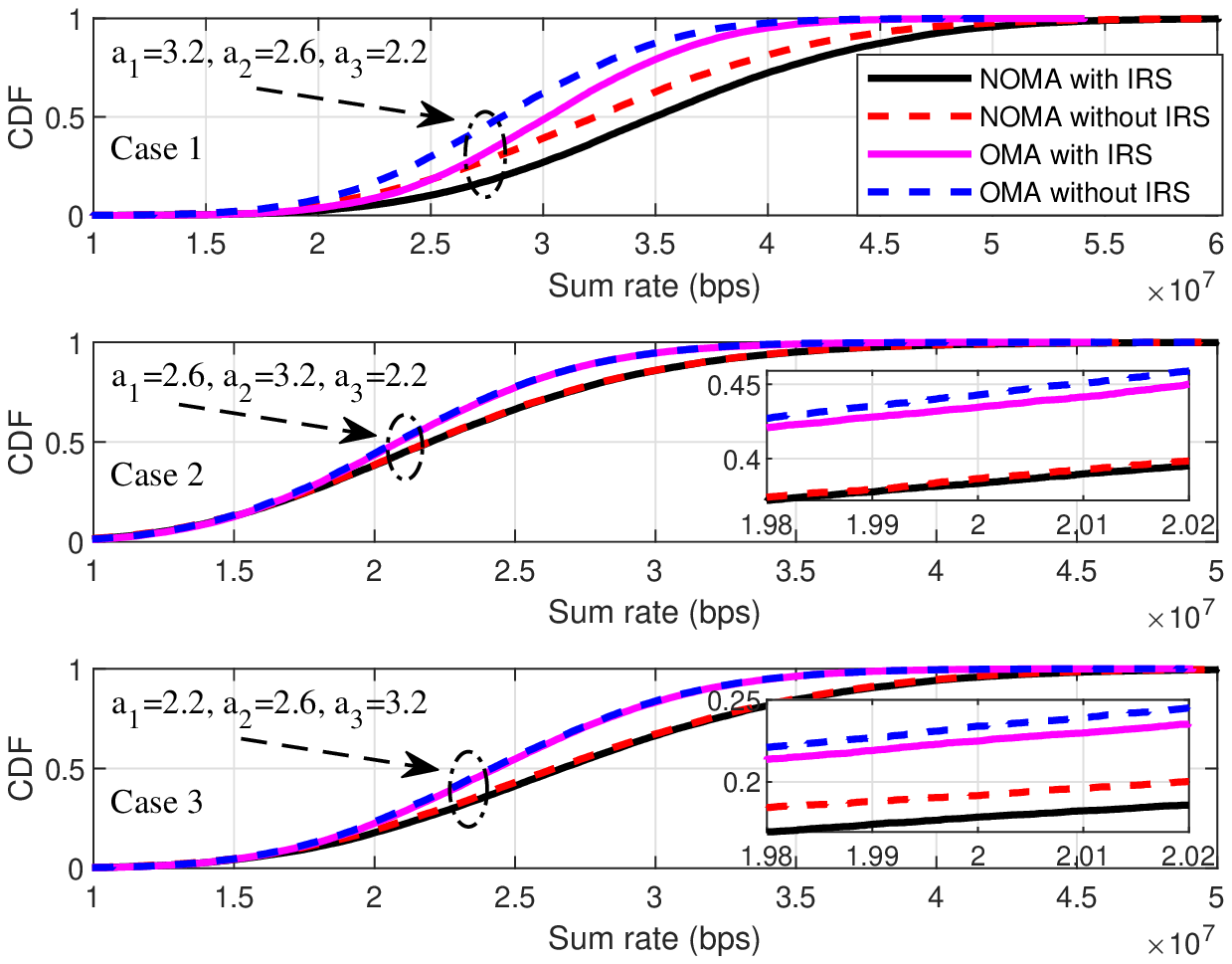}
		\caption{CDF under different settings.}
		\label{CDF_under_different_attenuation}
	\end{minipage}
	\hspace{2 mm}
	\begin{minipage}[t]{0.3 \textwidth}
		\centering
		\includegraphics[width=2.3 in]{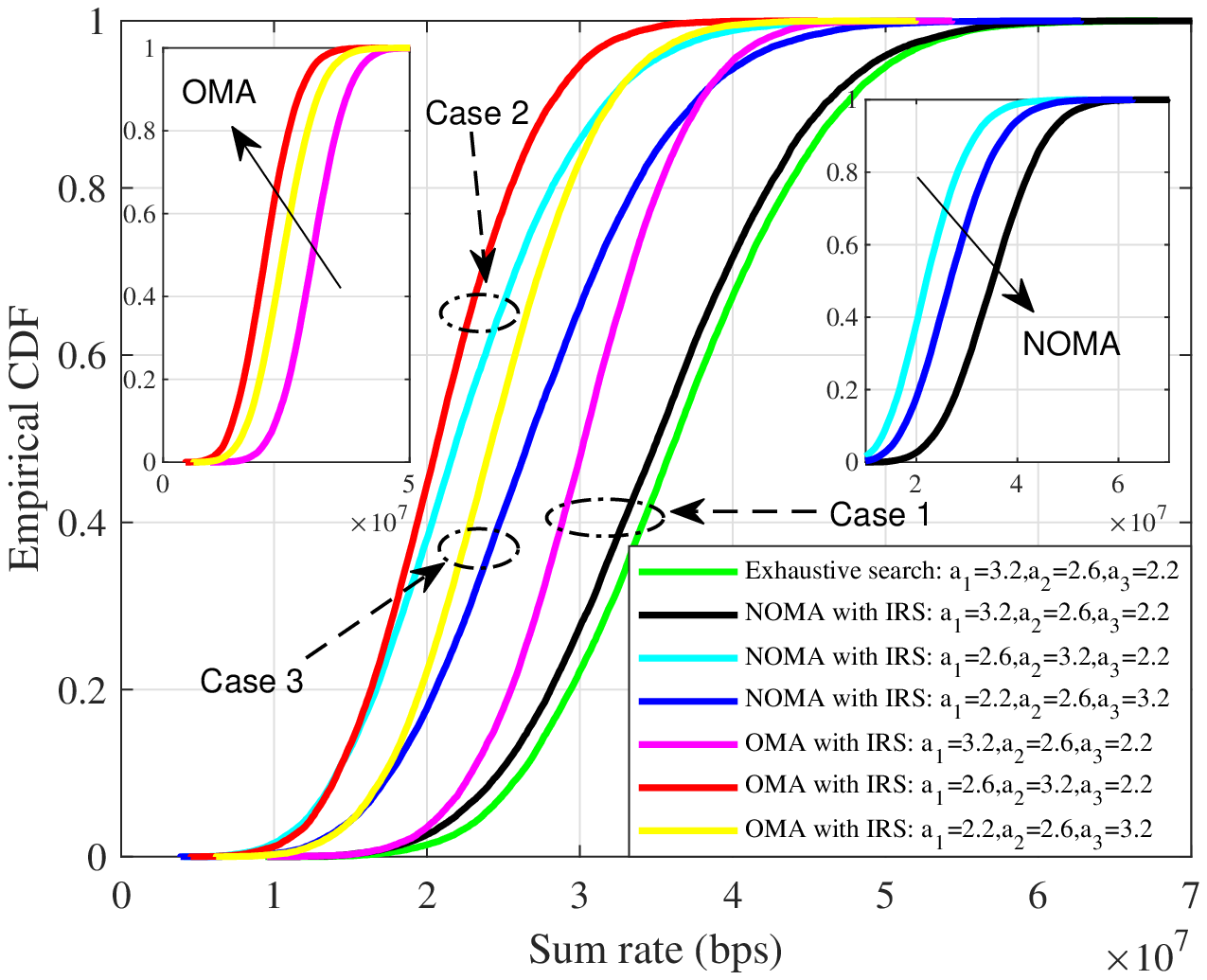}
		\caption{Empirical CDF of sum rate.}
		\label{ES_CDF_sum_rate}
	\end{minipage}
\end{figure*}

\subsection{Performance Analysis of Achievable Sum Rate} \label{sum_rate}
{Fig. \ref{ES_individual_data_rate} demonstrates the individual data rate of all users, where the exhaustive search is invoked to solve problems w.r.t. user association, subchannel assignment, and decoding order.
It can be observed that the data rate of user 3 and user 5 is much lower than that of other users.
This is mainly because that the objective of this paper is to maximize the sum rate of all users under the case of equal weights, which inevitably ignores user fairness.
Optionally, unequal weights can be used for controlling the fairness among users.
Namely, by assigning higher weights to the users having poor channel conditions, a more balanced data rate distribution can be achieved.
Note that with the aid of the algorithms designed in this paper, the resource allocation problems in the case of unequal weights can be easily solved without much effort.

Fig \ref{CDF_under_different_attenuation} evaluates the CDF performance of our proposed algorithms and benchmarks under different settings of path loss exponents.
As we can see, the IRS in case 1 (i.e., $a_{1}=3.2, a_{2}=2.6, a_{3}=2.2$) is capable of providing the most significant performance gain compared to the other two cases, where the path loss exponent of the BS-user, IRS-user and BS-IRS links are denoted by $a_{1}$, $a_{2}$ and $a_{3}$, respectively.
This is due to the fact that the path loss exponents of IRS-related links in the case 2 and case 3 are much larger than that of case 1, which worsens the signal attenuation of the reflective links and makes the performance gains vanishing.
As a result, one can know that if the benefits of reflective links brought by tuning the phase shifts of IRS are suppressed by the unfavorable large-scale fading, the system throughput cannot be effectively improved even when large intelligent surfaces are deployed.

%
Fig. \ref{ES_CDF_sum_rate} illustrates the effectiveness and sub-optimality of our proposed algorithms.
On the one hand, it can be seen at the probability of 60\% that the curves of case 1 are closer to the right side than the other two cases, and case 3 performs better than case 2.
Thereby, the solutions in case 1 are champions, the schemes in case 3 are runners-up, and the settings in case 2 are third-places.
On the other hand, it can be observed from the curves in the same case that NOMA schemes enjoy a significant performance gain than OMA counterparts.
This is because NOMA allows multiple users to simultaneously reuse the same subchannel, and thus obtains a higher spectrum efficiency.
In addition, it is well known that the exhaustive search always outperforms the non-exhaustive algorithms at the cost of complexity.
However, our proposed algorithm can achieve close performance to that obtained by the exhaustive search.
To be specific, when $a_{1}=3.2$, $a_{2}=2.6$ and $a_{3}=2.2$, the designed algorithm for IRS-aided NOMA networks can obtain around $96.4\%$ of the system throughput achieved by the exhaustive search.}

\begin{figure*}[t]
	\centering
	\begin{minipage}[t]{0.48 \textwidth}
	\centering
	\includegraphics[width=2.8 in]{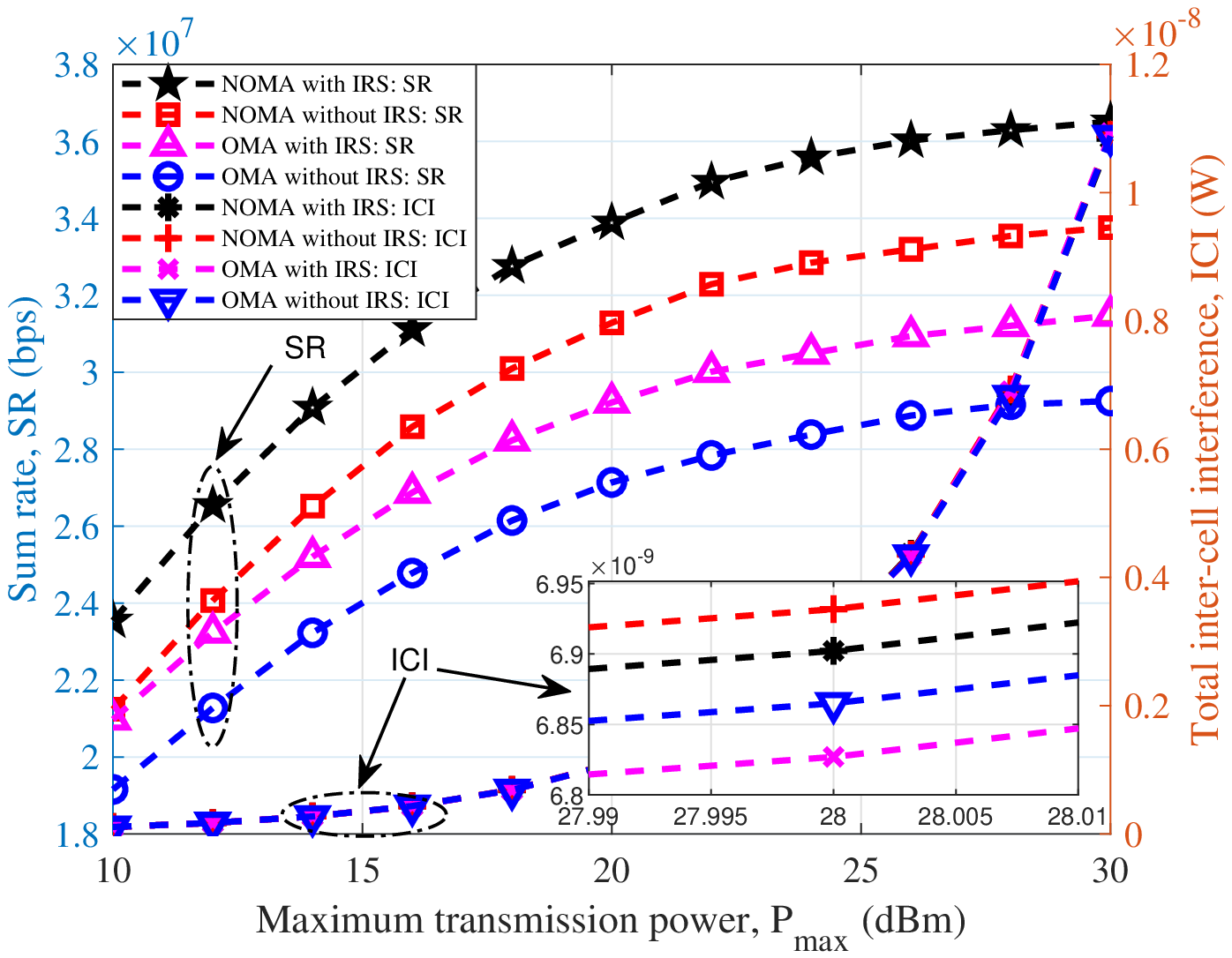}
	\caption{SR and ICI vs. maximum transmission power.}
	\label{sum_rate_vs_interference}
	\end{minipage}
	\begin{minipage}[t]{0.45 \textwidth}
		\centering
		\includegraphics[width=2.8 in]{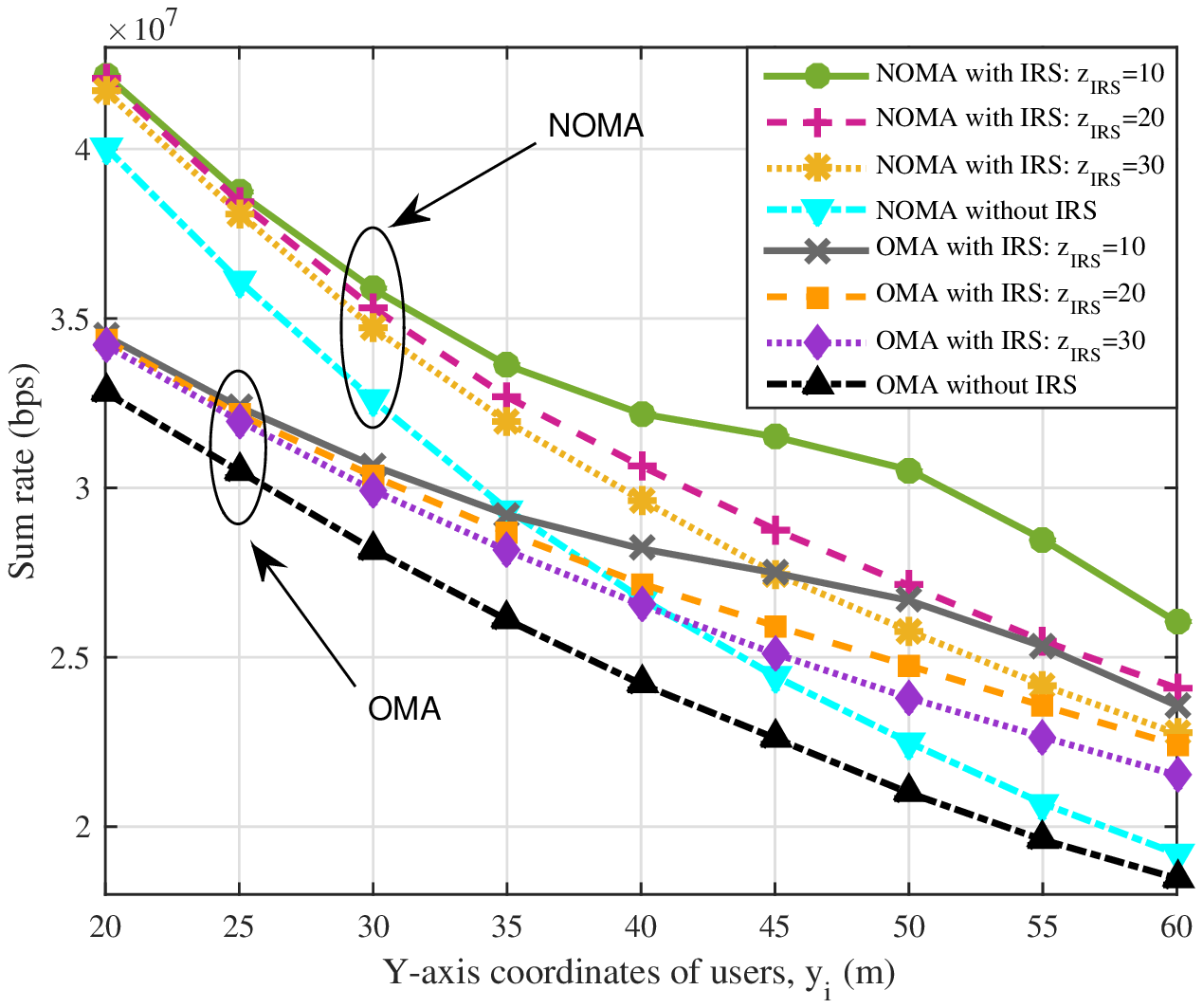}
		\caption{Sum rate vs. user location.}
		\label{sum_rates_versus_user_location}
	\end{minipage}
\end{figure*}

Fig. \ref{sum_rate_vs_interference} compares the impact of the maximum transmission power on the achievable sum rate.
Similar observations are achieved in Fig. \ref{CDF_under_different_attenuation} and Fig. \ref{ES_CDF_sum_rate}, e.g., NOMA schemes outperform their OMA counterparts, and the performance gain becomes more significant when IRS is leveraged.
It can be noticed that the lower the $P_{\text{max}}$ value is, the larger slope of the sum rate curves is.
Thus, different from the approximately linear growth at a low $P_{\text{max}}$, the sum rate curves increase more slowly at a high $P_{\text{max}}$ due to the existence of intra-cell and inter-cell interference.
It’s worth pointing that the performance of NOMA/OMA schemes would reach their peak as the maximum transmission power increases to a certain threshold, and more reflecting elements have to be equipped on the IRS to further eliminate interference and improve performance.

Fig. \ref{sum_rates_versus_user_location} characterizes the achievable sum rate versus the locations of all users in diverse situations.
When the coordinates of BSs and IRS are fixed, moving users farther away from all BSs by increasing their y-axis coordinates will lead to a lower sum rate.
In this case, although users will be closer to the IRS, the increased signal attenuation is the dominant factor compared to power gain provided by the IRS.
Specifically, it is worth noting that when $y_{i} \ge 40$ the OMA schemes with IRS outperform the conventional NOMA without IRS, this is because the power gain compensated by the IRS greater than the performance gap between OMA and NOMA.
More particularly, when $y_{i} = 60$ and $z_{\text{IRS}}=10$, it can be obtained that the multi-cell IRS-aided NOMA and OMA networks are capable of providing up to 35.4\% and 22.7\% higher sum rate than the conventional NOMA and OMA schemes, respectively.

\subsection{Performance Analysis of System Energy Efficiency} \label{EE}
\begin{figure*}[t!]
	\begin{minipage}[t]{0.3 \textwidth}
		\centering
		\includegraphics[width=2.3 in]{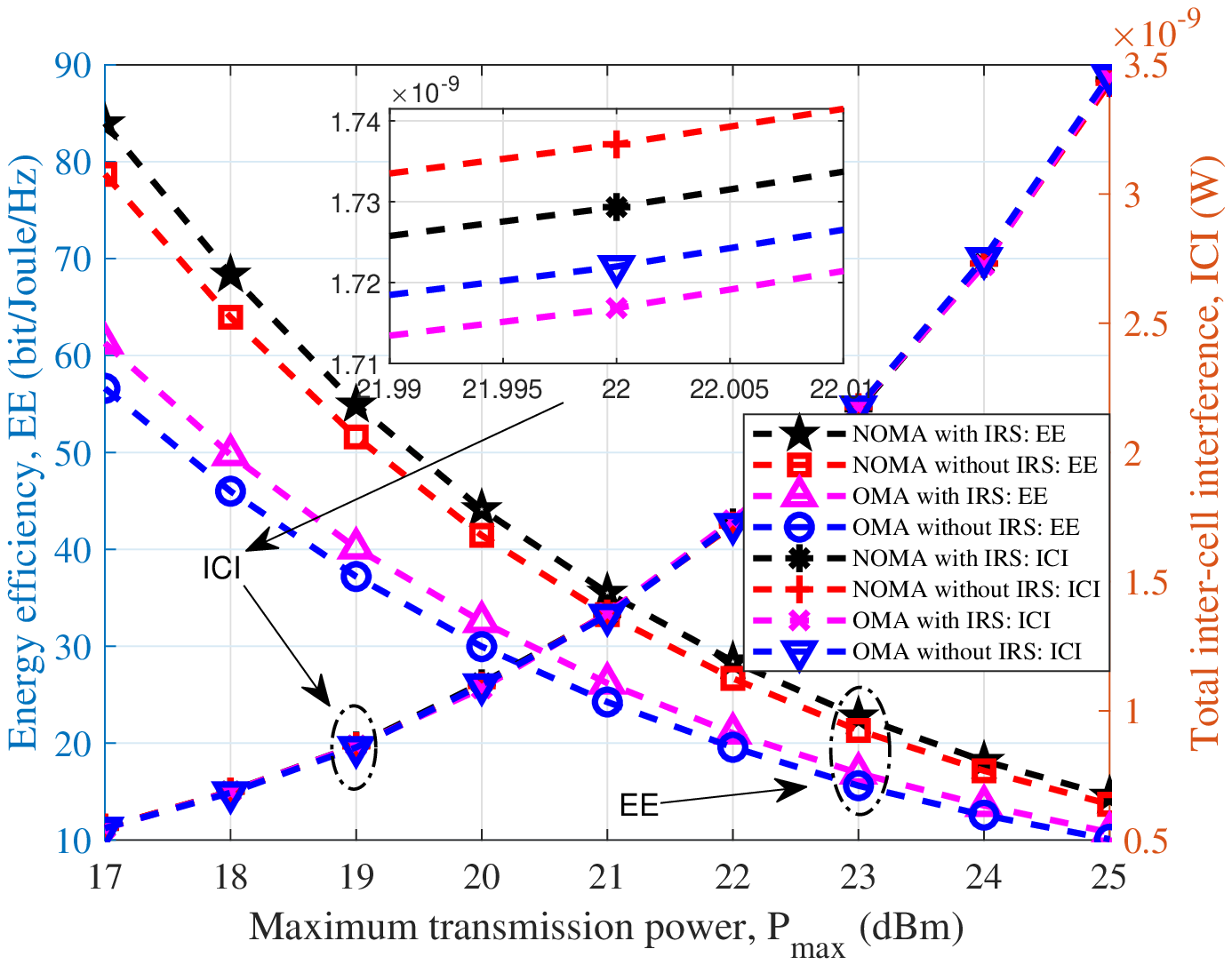}
		\caption{EE and ICI vs. power budget.}
		\label{energy_efficiency_vs_interference}
	\end{minipage}
	\hspace{6 mm}
	\begin{minipage}[t]{0.3 \textwidth}
		\centering
		\includegraphics[width=2.3 in]{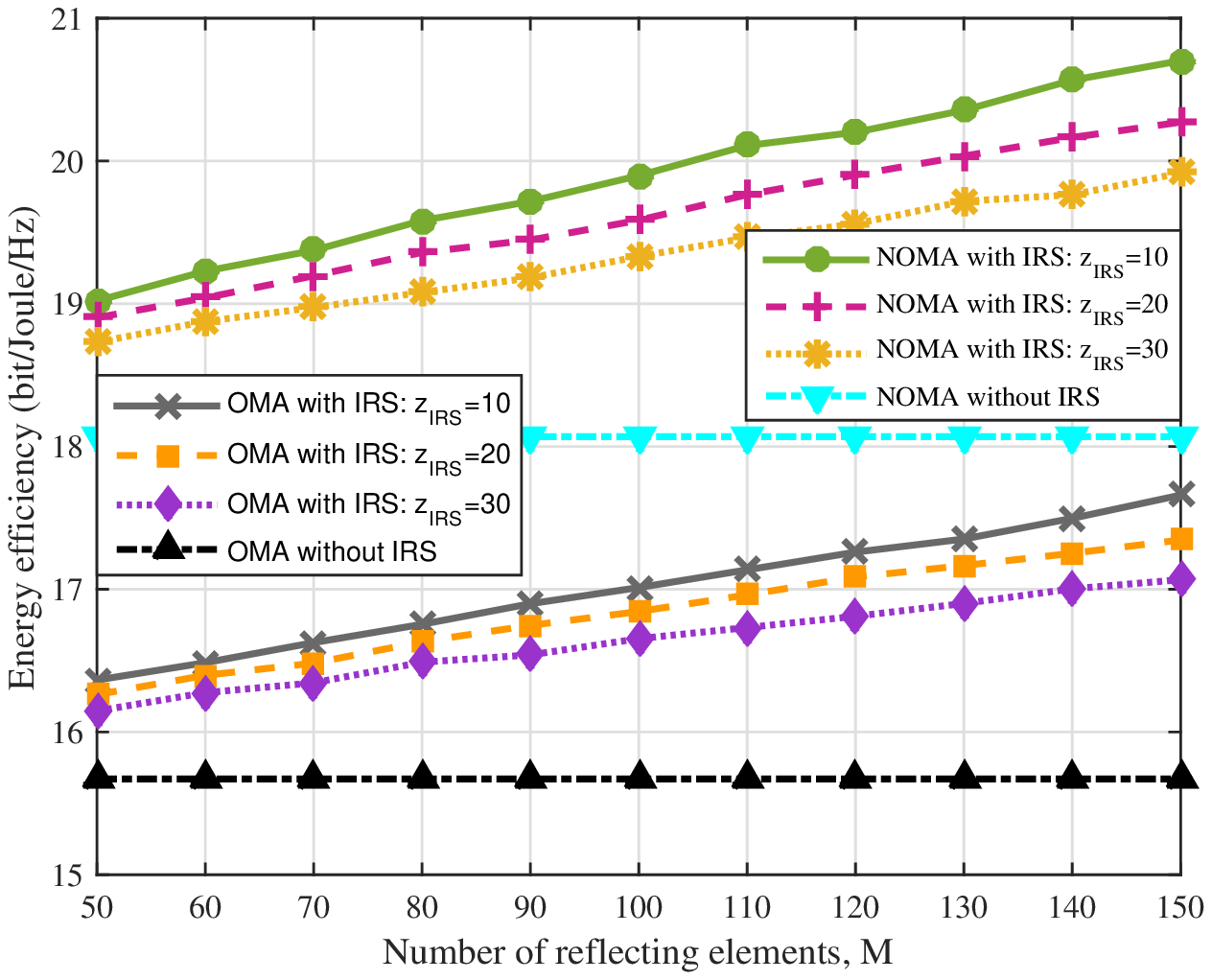}
		\caption{EE vs. reflecting elements.}
		\label{energy_efficiency_versus_M}
	\end{minipage}
	\hspace{2 mm}
	\begin{minipage}[t]{0.3 \textwidth}
		\centering
		\includegraphics[width=2.3 in]{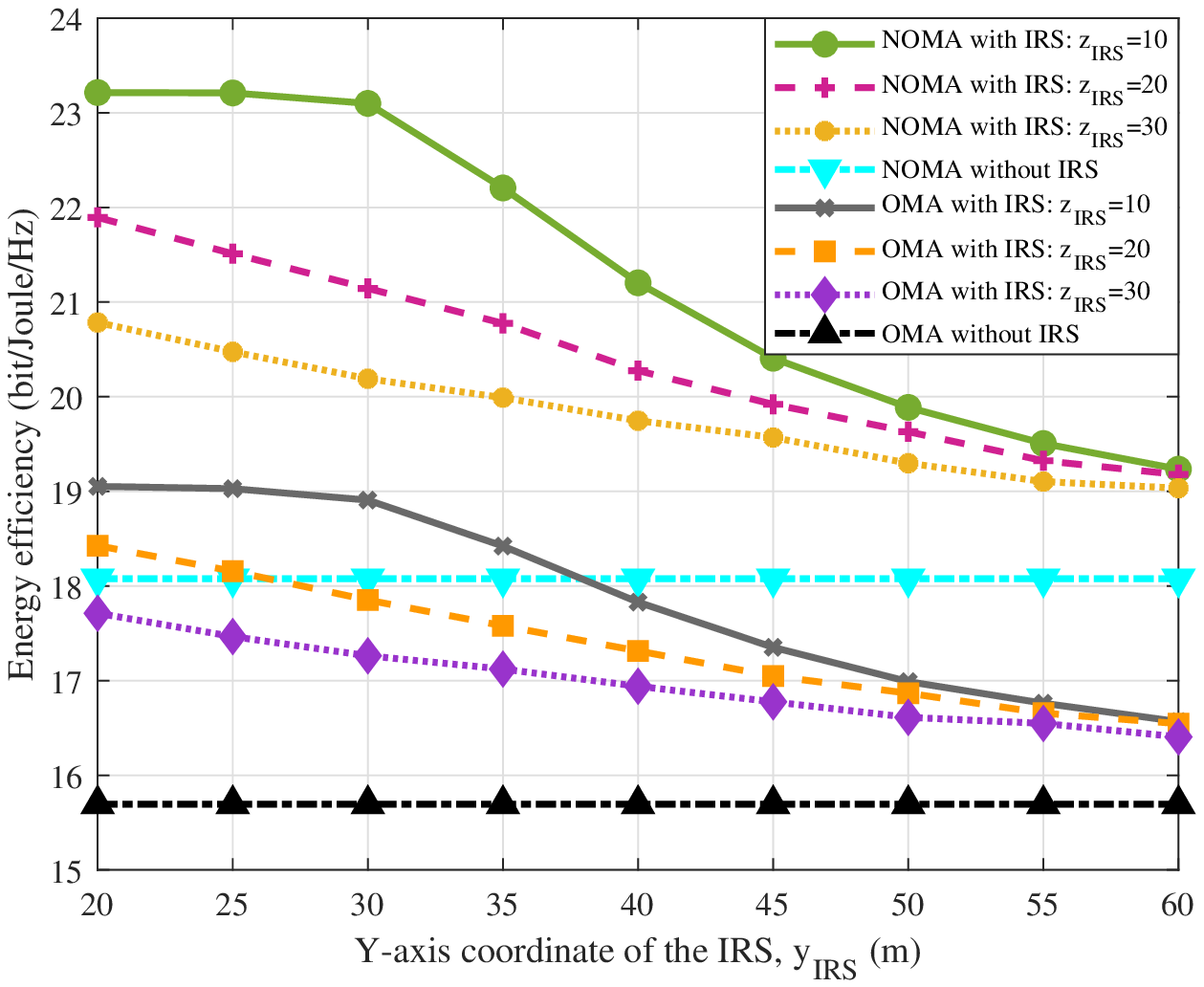}
		\caption{EE vs. the location of IRS.}
		\label{energy_efficiency_versus_IRS_location}
	\end{minipage}
\end{figure*}

{Fig. \ref{energy_efficiency_vs_interference} shows the impact of the maximum transmission power on energy efficiency.
On the one hand, it can be observed that the trend in this figure is opposite to that in Fig. \ref{sum_rate_vs_interference}, where the energy efficiency decreases as the maximum transmission power increases.
The reason is that the objective of maximizing sum-rate requires all available power at the BS, which is different to energy efficiency maximization.
Hence, the proposed algorithms for sum-rate maximization objective lead to the decrease of energy efficiency.
Indeed, once the QoS constraints are met, the energy efficiency becomes better when using full transmission power for a lower $P_{\text{max}}$.
This is due to the fact that the interferences experienced by users are weak for low $P_{\text{max}}$, and the available transmission power can be fully utilized.
In contrast, for large $P_{\text{max}}$, the increased interferences deteriorate the energy efficiency rapidly.
On the other hand, it shows that the larger the $P_{\text{max}}$ value is, the lower the slope of the energy efficiency curves will be.
This is due to the fact that less performance gain will be obtained in terms of sum rate, when the BS's maximum transmission power continues to increase.
Moreover, it can be noticed that NOMA schemes experience higher inter-cell interference than their OMA counterparts due to the severe reuse of the same time-frequency resource among multiple cells.
Thanks to the deployment of IRS, one can also observe that the inter-cell interference is able to be effectively eliminated, the energy efficiency can be thus further improved.}


Fig. \ref{energy_efficiency_versus_M} plots the energy efficiency versus the number of reflecting elements at different heights.
The trend can be sketched that the larger the $M$ value is, the larger energy efficiency will be obtained in the IRS-aided networks.
Furthermore, compared to the benchmark schemes without IRS, the performance gains of energy efficiency in the IRS-aided NOMA networks are larger than that in the OMA schemes with IRS.
This is because IRS can be used to suspend interferences and enhance desired signals at the same time in the NOMA networks, while it can only play a role in enhancing the signals in the considered OMA schemes.
%
Particularly, when $M=140$ and $z_{\text{IRS}}=10$, it can be obtained that the IRS-aided NOMA and OMA networks are capable of enjoying 13.8\% and 11.6\% higher energy efficiency than the conventional NOMA and OMA schemes, respectively.
Thus, it is a direct consequence of the fact that better performance can be achieved by employing a large number of reflecting elements to alleviate interferences and enhance the desired signals in the multi-cell IRS-aided NOMA networks.

Fig. \ref{energy_efficiency_versus_IRS_location} demonstrates the impact of the location of the IRS on energy efficiency,
{where the results are averaged 10,000 trials.}
As it can be observed, when the locations of all BSs and users are fixed, the increase of y-axis coordinate of IRS leads to the degradation of the energy efficiency.
This is due to the fact that the larger the BS-IRS and IRS-user distances are, the larger path loss will be experienced by the reflective channel, and the smaller power gain will be brought by the IRS.
Besides, when the height of the IRS drops, there is a slight performance improvement at the cost of coverage.
In other words, there exists a trade-off between the sum rate and coverage area when we integrate IRS into wireless networks.
Similar observations are achieved in Fig. \ref{energy_efficiency_versus_M}, the IRS allows the available power in the NOMA-based networks to be used more efficiently.
{Concretely, when $y_{\text{IRS}}=35$ and $z_{\text{IRS}}=10$, it can be found that IRS-aided NOMA/OMA networks are capable of enjoying 22.8\% and 12\% higher energy efficiency than conventional NOMA/OMA counterparts, respectively.}
\end{spacing}

\begin{spacing}{1.4}
\section{Conclusions}
\label{conclusion}
In this paper, we investigated the sum-rate maximization problem in the IRS-aided multi-cell NOMA network, which was formulated as a MINLP problem.
Then, relaxation methods were invoked to transform the intractable subproblems into convex ones, and efficient algorithms were designed to solve these challenging subproblems iteratively.
Next, in order to achieve a two-sided exchange-stable state among users, BSs and subchannels, swap matching-based algorithms were proposed.
Finally, numerical results under various settings demonstrated that through proactively reconfiguring the wireless communication environment,
the IRS is capable of enhancing the system performance.
Additionally, the proposed algorithms can significantly improve both the system throughput and energy efficiency.


{
\appendices
\section{} \label{proof_of_remark_1}
Based on the complexity analysis in Section \ref{section_3_analysis} and Section \ref{section_4_analysis}, the computational complexity of \textbf{Step 1-2-3-4} can be given as $\mathcal{O}_{1}$, $\mathcal{O}_{2}$, $\mathcal{O}_{3}$, and $\mathcal{O}_{4}$, respectively.
Then, the complexity of \textbf{Algorithm \ref{alternating_algorithm}} can be represented as
$\bar{\mathcal{O}} = \big( \mathcal{O}_{0} + N_{1} \left( \mathcal{O}_{1} + \mathcal{O}_{2} + \mathcal{O}_{3} + \mathcal{O}_{4} \right) \big)$,
where
$\mathcal{O}_{0} = N_3 (2IJK)^3$ denotes the complexity of \textbf{Algorithm \ref{algorithm_feasibility_searching}},
$\mathcal{O}_{1} = N_2 (2IJK)^3$,
$\mathcal{O}_{2} = (M+4IJK)^6 + N_4 T_{GR}$,
$\mathcal{O}_{3} = IJ^2 + A_{\text{max}}IJN_{it}$,
and $\mathcal{O}_{4} = J^2 K^2(\bar{N}_{it}+1)$.

According to the convergence analysis in Section \ref{section_3_analysis} and Section \ref{section_4_analysis}, one can know that the objective value of problem (\ref{max_sum_rate}) in \textbf{Algorithm \ref{alternating_algorithm}} is non-decreasing over iterations.
Moreover, the system throughput is upper bounded due to the limited bandwidth and power budget.
Therefore, the proposed \textbf{Algorithm \ref{alternating_algorithm}} is guaranteed to converge as long as $N_{1}$ is set large enough, which completes the proof of \textbf{Remark \ref{complexity_and_convergence}}.}
\end{spacing}



\begin{spacing}{1.05}
\bibliographystyle{IEEEtran}
\bibliography{IEEEabrv,ref}
\end{spacing}
\end{document}